\documentclass[twocolumn,trackchanges]{aastex62} 
\usepackage{verbatim, units}

\newcommand\hi{H{\small I}}
\newcommand\htoo{H$_2$}
\newcommand\nhtoo{$n_{\mathrm H_2}$}

\newcommand\tb{T$_\mathrm{B}$}
\newcommand\tkin{T$_{kin}$}

\newcommand\oiiihb{[\ion{O}{3}]/H$\beta$}

\newcommand\thirco{$^{13}$CO}
\newcommand\tweco{$^{12}$CO}
\newcommand\ceighto{C$^{18}$O}
\newcommand\eighto{$^{18}$O}

\newcommand\thirc{$^{13}$C}
\newcommand\twec{$^{12}$C}
\newcommand\hcop{HCO$^+$}
\newcommand\chhhoh{CH$_3$OH}

\newcommand\atlas{ATLAS$^{\rm 3D}$}
\newcommand\kms{km s$^{-1}$}
\def\arcsec{$^{\prime\prime}$}

\newcommand\solmass{M$_\odot$}

\newcommand\mjb{mJy~beam$^{-1}$}
\newcommand\jybkms{Jy bm$^{-1}$ km s$^{-1}$}
\newcommand\jykms{Jy~km~s$^{-1}$}
\newcommand\persqcm{~cm$^{-2}$}
\newcommand\percc{~cm$^{-3}$}
\newcommand\persqpc{~pc$^{-2}$}
\newcommand\peryr{~yr$^{-1}$}
\newcommand\been{\begin{enumerate}}
\newcommand\een{\end{enumerate}}
\newcommand\e[1]{$\times 10^{#1}$}
\newcommand\jthreetwo{$J=\frac{3}{2}-\frac{1}{2}$}  
\newcommand\jonetwo{$J=\frac{1}{2}-\frac{1}{2}$}  
\newcommand\msunpckkms{\solmass\persqpc~(K~\kms)$^{-1}$}
\newcommand\alphaCO{$\alpha_\mathrm{CO}$}

\received{17 Mar 2022}
\revised{06 May 2022}
\accepted{17 May 2022}

\shorttitle{Molecular properties of NGC 4526}
\shortauthors{Young et al.}

\begin{document}

\title{Down but not out: properties of the molecular gas in the stripped Virgo Cluster early-type galaxy NGC~4526}

\correspondingauthor{Lisa Young}
\email{lisa.young@nmt.edu}

\author[0000-0002-5669-5038]{Lisa M.\ Young}
\affil{Physics Department, New Mexico Tech,
801 Leroy Place,
Socorro, NM 87801, USA}
\affil{Adjunct Astronomer, National Radio Astronomy Observatory, Socorro, NM 87801, USA}
\author[0000-0001-9436-9471]{David S.\ Meier}
\affil{Physics Department, New Mexico Tech,
801 Leroy Place,
Socorro, NM 87801, USA}
\affil{Adjunct Astronomer, National Radio Astronomy Observatory, Socorro, NM 87801, USA}
\author[0000-0001-8513-4945]{Alison Crocker}
\affil{Department of Physics, Reed College, Portland, OR 97202, USA}
\author[0000-0003-4932-9379]{Timothy A.\ Davis}
\affil{School of Physics \& Astronomy, Cardiff University, Queens Buildings, The Parade, Cardiff, CF24 3AA, UK}
\author[0000-0003-2132-5632]{Sel{\c c}uk Topal}
\affil{Department of Physics, Van Y\"uz\"unc\"u Y{\i}l University, Van 65080, Turkey}


\begin{abstract}

We present ALMA data on the 3mm continuum emission, CO isotopologues (\tweco, \thirco, and \ceighto), and high-density molecular tracers (HCN, \hcop, HNC, HNCO, CS, CN, and \chhhoh) in NGC~4526.
These data enable a detailed study of the physical properties of the molecular gas in a longtime resident of the Virgo Cluster; comparisons to more commonly-studied spiral galaxies offer intriguing hints into the processing of molecular gas in the cluster environment.
Many molecular line ratios in NGC~4526, along with our inferred abundances and CO/\htoo\ conversion factors, are similar to those found in nearby spirals.
One striking exception is the very low observed \tweco/\thirco(1$-$0) line ratio, $3.4\pm0.3$, which is unusually low for spirals though not for Virgo Cluster early-type galaxies. 
We carry out radiative transfer modeling of the CO isotopologues with some archival (2$-$1) data, and we
use Bayesian analysis with Markov chain Monte Carlo techniques to infer the physical properties of the CO-emitting gas.  We find surprisingly low [\tweco/\thirco] abundance ratios of $7.8^{+2.7}_{-1.5}$
and $6.5^{+3.0}_{-1.3}$ at radii of 0.4 kpc and 1 kpc.
The emission from the high-density tracers HCN, \hcop, HNC, CS and CN is also relatively bright, and CN is unusually optically thick in the inner parts of NGC~4526.  These features hint that processing in the cluster environment may have removed much of the galaxy's relatively diffuse, optically thinner molecular gas along with its atomic gas.  Angular momentum transfer to the surrounding intracluster medium may also have caused contraction of the disk, magnifying radial gradients such as we find in [\thirco/\ceighto].  More detailed chemical evolution modeling would be interesting in order to explore whether the unusual [\tweco/\thirco] abundance ratio is entirely an environmental effect or whether it also reflects the relatively old stellar population in this early-type galaxy.

\end{abstract}


\section{Introduction} \label{sec:intro}

One of the outstanding questions in modern astrophysics is understanding the origins of the Hubble sequence and the surprising diversity of nearby galaxies.  Fortunately, nearby galaxies preserve some clues to their histories in their current properties. Cold molecular gas in early-type (elliptical and lenticular) galaxies is particularly interesting in this context; it falls outside our simple paradigm that early-type galaxies are quiescent and free of cold gas.
In fact, almost 25\% of nearby early-type galaxies have retained some molecular gas and at least 40\% of them host atomic and/or molecular gas \citep{a3d_cmd,davis_massiv} at a level of $\gtrsim 10^{-3}$ M$_\star$.

Detailed studies of the gas in early-type galaxies have provided important insights into their evolutionary histories.
Comparisons between gas and stellar kinematics reveal frequent kinematic misalignments between the gas and stars, such that it is not unusual for the cold gas in early-type galaxies to be counterrotating with respect to the stars.  That counterrotating gas cannot have had a long symbiotic relationship with those stars.
Up to 30\% -- 50\% of the cold gas in early-type galaxies is sufficiently misaligned that it must have come in from outside after the bulk of the stars were formed \citep{davis_misalign}.
More recently, \citet{davis+young} moved towards using metallicities as another, complementary set of clues to the evolution of early-type galaxies.  Signatures of gas accretion can be seen in the fact that the metallicity of the ionized gas in early-type galaxies is occasionally lower than the metallicity of the stars.
Ultimately it would be useful also to study the isotopic abundance patterns in early-type galaxies as further clues to their gas accretion and nucleosynthetic enrichment histories, employing the types of chemical evolution models that are frequently used for the Milky Way \citep{romano2019,nupycee}.

Beyond their accretion histories, early-type galaxies also allow us to probe the processing of galaxies in clusters.  And because they have generally been resident in the clusters for a much longer time than the spirals that are more commonly the subject of ram-pressure stripping studies, they give additional insights beyond studies of spirals.  They sample longer timelines for the cluster-driven processing.

In this paper we present an exploration of molecular line ratios in the Virgo Cluster galaxy NGC~4526.  The target is an unusually molecule-rich early-type galaxy with a striking but small silhouette dust disk, 1 kpc in radius.
We focus on new ALMA data for continuum emission, CO isotopologues, and high-density molecular tracers in the 3mm band, supplemented by archival \tweco(2$-$1) and \thirco(2$-$1) data.
We carry out radiative transfer modeling of the available isotopologue data, coupled with Bayesian inference techniques.  The current data constrain the excitation-corrected [\tweco/\thirco] and 
[\thirco/\ceighto] abundance ratios; we find the former to be roughly constant in the disk but the latter to have a strong radial gradient, and those patterns provide clues to the processes driving the abundance gradients.  We also estimate the CO/\htoo\ conversion factors \alphaCO\  using the dust continuum and CO images, providing the first estimate of this type for an early-type galaxy.
This analysis 
lays the groundwork for future explorations of isotopic abundances in early-type galaxies, and especially for cluster members, which should give insights into the evolution of early-type galaxies and the processing of galaxies in clusters.

\section{About ngc~4526}\label{sec:about}

\begin{figure*}
\includegraphics[width=\textwidth, trim=2mm 2mm 0cm 0cm, clip]{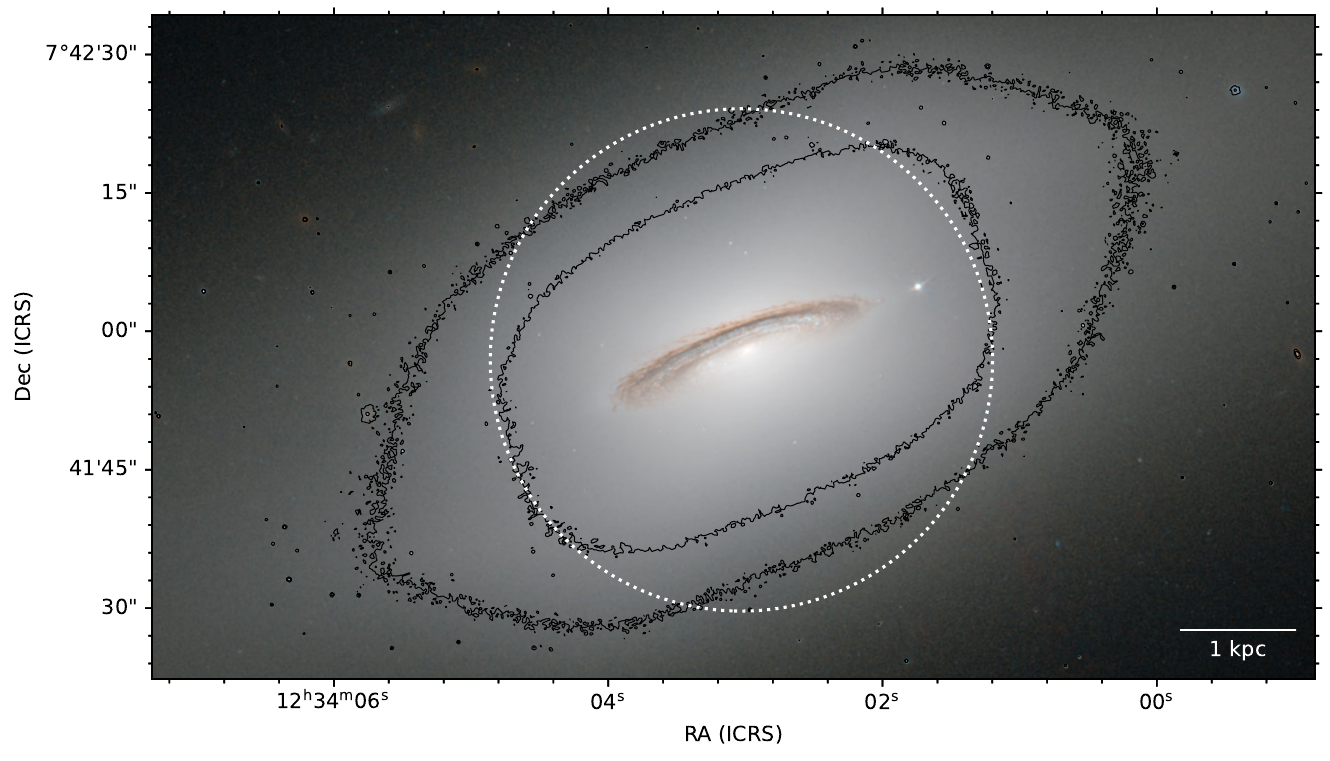}
\caption{NGC~4526.  The white circle shows the FWHM of the ALMA field of view at 110 GHz.  The background color image is HST data, provided by ESA/Hubble and NASA (https://esahubble.org/images/potw1442a/); original image by Judy Schmidt, with astrometry by LY.  Two optical contours are overlaid to highlight the boxy structure in the galaxy's bulge.\label{fig:4526_hst}}
\end{figure*}

NGC~4526 is a Virgo Cluster member whose prominent central dust disk (Figure \ref{fig:4526_hst}) has long prompted interest in its cold gas content.  
CO emission was detected in the galaxy in single dish surveys by \citet{sage&wrobel} and \citet{CYB}; the first resolved images of its CO emission were published by \citet{YBC_S0s}.
It is also a member of the \atlas\ survey \citep{cappellari_a3d1}, which provides additional information on its stellar and ionized gas kinematics \citep{davor}, stellar populations \citep{mcdermid2015}, and star formation history \citep{davor_ageZ}.
It is a fast rotator and
one of the more massive galaxies in the \atlas\ sample, with a stellar mass $\log (M_\star/M_\odot) \sim 11.1$ and global colors that place it firmly in the red sequence \citep{a3d_cmd}.
As a member of the Virgo Cluster, we assume a distance of 16.4 Mpc \citep{cappellari_a3d1}.
A more recent measurement based on the tip of the red giant branch gives 15.7 $\pm$ 0.2 $\pm$ 0.4 Mpc \citep[the uncertainties are statistical and systematic, respectively;][]{hatt2018}, which is only 6\% smaller than the distance we have assumed.

\citet{davis_misalign} studied the kinematics of gas and stars in the \atlas\ sample and noted that field early-type galaxies often display kinematic misalignments between their gas and stars, suggesting the gas was recently accreted from some external source.  In contrast, the Virgo Cluster early-type galaxies and those in other dense groups tend to have relaxed and well-aligned prograde gas.  In this respect NGC~4526 is typical of cluster members.

It would have been difficult for NGC~4526 to acquire gas after falling into the Virgo Cluster, because the relative velocities of typical interactions (characterized by the velocity dispersion of the cluster) are much larger than the internal escape velocities of the galaxies themselves.
Thus, unlike the cold gas in many field early-type galaxies, this gas has most likely been in NGC~4526 for several Gyr and has been processed through a long, continuous symbiotic relationship with the stars in multiple generations of mass loss and star formation.

The molecular gas in NGC~4526 is also undergoing star formation with an efficiency not too different from that of typical spirals.
\citet{davis_sfr} estimate a star formation rate of 0.2 \solmass\peryr\ based on 22 \micron\ and FUV emission; this gives a specific star formation rate (SFR/M$_\star$) of $10^{-11.8}$\peryr, a gas depletion time of 2.6 Gyr, and a star formation rate surface density $\Sigma_{\rm SFR} \sim 0.09$ \solmass\peryr~kpc$^{-2}$.  Those values are all averaged over the whole molecular disk.

Previous observations of the cold gas in the galaxy have revealed some unusual and extreme properties.
For example, \citet{crocker_hd} used the IRAM 30m telescope to study molecular tracers in a subset of the \atlas\ early-type galaxies.
They found that, relative to \tweco, NGC~4526 has unusually bright \thirco\ and HCN and a high HCN/\hcop\ line ratio.
It is also undetected in \hi\ emission \citep{lucero+young}, which produces a remarkably large \htoo/\hi\ mass ratio $>$ 60. 
This extreme deficit of atomic gas suggests that the low-density ISM in NGC~4526 has been stripped by the intracluster medium.  Its molecular properties should therefore give valuable perspective on the effects of a long residence in a cluster.

\section{Observations}

NGC~4526 was observed in ALMA's compact configurations 
in band 3, projects 2017.1.01108.S and 2018.1.01599.S, in April 2018 and April 2019.
We made use of the standard pipeline calibrated raw data and carried out continuum subtraction, imaging, and cleaning ourselves.  An additional round of phase selfcal was not found to offer any improvement in the images because of the relatively modest signal-to-noise ratios.
For continuum subtraction we used zero-order or first-order fits to the line-free channels in the visibility domain; for imaging, we made a wide variety of images at varying channel widths and resolutions as necessary to optimize the resolution or to match resolutions for resolved line ratios. The times on source, along with fiducial beam sizes and rms noise levels, are indicated in Table \ref{tab:rms} for the detected spectral lines.  Emission was cleaned down to about the 1$\sigma$ level and a primary beam correction was applied.

We also use \tweco(2-1) observations obtained with the CARMA array \citep{davis_nature}.  Those observations are relatively high resolution data, and were initially imaged at 0\farcs25 resolution; for the analysis here, we tapered the visibility data to make a lower-resolution image and smoothed in the image plane to match the resolution of the ALMA \tweco(1-0) data.  Appendix \ref{carmadata} describes short-spacing verification and continuum subtraction on the \tweco(2-1) data.

\begin{deluxetable}{lllcc}
\tablecaption{Fiducial image parameters for NGC~4526\label{tab:rms}}
\tablehead{
\colhead{Line} & \colhead{Time} &
\colhead{Beam size} &
\colhead{rms} & \colhead{$\Delta v$}\\
\colhead{} & \colhead{sec} & \colhead{(\arcsec)} &
\colhead{mJy bm$^{-1}$} & \colhead{\kms}
}
\startdata
\tweco(1$-$0) & 1727 & 2.01$\times$1.90 & 0.97  & 15.0 \\
\tweco(1$-$0) & 1727 & 1.12$\times$0.82 & 2.0  & 2.54 \\
\tweco(2$-$1) & \nodata\tablenotemark{a} & 1.12$\times$1.12 & 9.0 & 10.0 \\ 
CN (N=1$-$0)\tablenotemark{b}  & 1727 & 1.47$\times$1.14 &  0.19 & 41.3\\ 
\thirco(1$-$0) & 5443 & 2.01$\times$1.89 & 0.36 & 15.0\\  
HNCO$(5_{0,5}-4_{0,4})$ & 5443 & 2.01$\times$1.90 & 0.36 & 15.0\\
\ceighto(1$-$0) & 5443 & 2.01$\times$1.90 & 0.36 & 15.0\\  
CS(2$-$1) & 5443 & 2.32$\times$2.12 & 0.28 & 15.0\\   
\chhhoh ($2_k-1_k$)\tablenotemark{c} & 5443 & 2.35$\times$2.15 & 0.29 & 15.0 \\                 
HNC(1$-$0) & 3871 & 2.55$\times$2.38 & 0.26 & 15.0\\
\hcop(1$-$0) & 3871 & 2.60$\times$2.40 & 0.28 & 15.0\\
HCN(1$-$0) & 3871 & 2.60$\times$2.41 & 0.29 & 15.0\\
HNCO$(4_{0,4}-3_{0,3})$ & 3871 & 2.62$\times$2.43 & 0.29 & 15.0\\
cont.\ 99.3 GHz &  & 2.00$\times$2.00 & 0.012 & \\
cont.\ 99.3 GHz &  & 0.91$\times$0.79 & 0.014 & \\
\enddata
\tablenotetext{a}{Approximately 50 hours at CARMA, in four array configurations \citep{davis_nature}.}
\tablenotetext{b}{Both J=3/2$-$1/2 and J=1/2$-$1/2.}
\tablenotetext{c}{Four constituent transitions are blended.}
\tablecomments{Narrower velocity resolution is available for all the cubes made with 15 \kms\ channels but the faintest lines require binning for improved signal-to-noise.  For simplicity, we sometimes refer to the HNCO transitions as (5$-$4) and (4$-$3). In most cases, these are the `natural'-weighted beam sizes.
}
\end{deluxetable}

\section{Continuum}\label{sec:cont}

In NGC~4526 we have near-simultaneous observations in the same telescope configuration, covering frequencies from 87 GHz to 110 GHz (with gaps) with similar uv coverage and time on source, so these data are suitable for imaging the continuum intensity and estimating the spectral index of any emission.
We used all of the available line-free frequencies in CASA's multi-term, multi-frequency synthesis deconvolver \citep{rau_mfs}, and we solved for two terms, i.e.\ the intensity and the spectral slope.
When imaged at a relatively high resolution of 0.9\arcsec $\times$0.8\arcsec, the resulting continuum image at 99.33 GHz shows only a nuclear point source of flux density 4.79 $\pm$ 0.03 mJy and spectral index $-1.07 \pm 0.02$.  This nuclear source is clearly synchrotron emission.
\citet{chandra_atlas} have commented that the galaxy was not previously known to host an AGN; however, they detect a nuclear X-ray point source, and combined with a 5 GHz point source \citep{nyland_nuclei} and the 99 GHz synchrotron emission, these data are all consistent with the presence of an AGN. 

\begin{figure}
\includegraphics[width=\columnwidth, trim=0.3cm 1.7cm 1.5cm 3cm, clip]{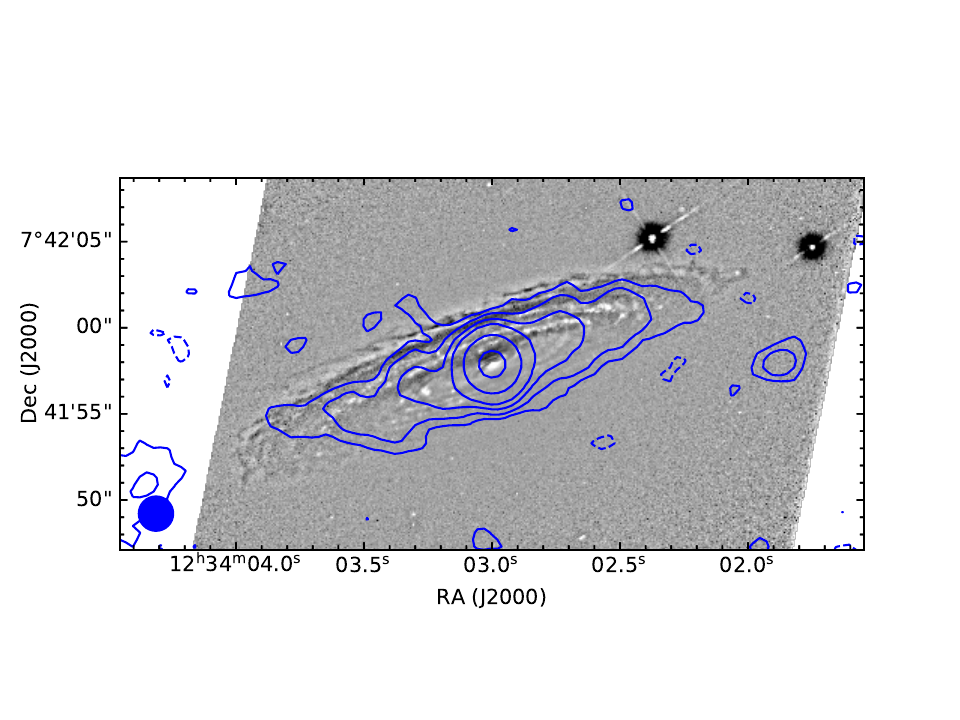}
\caption{3mm continuum emission.  The greyscale is an unsharp-masked HST WFPC2 image in the F555W filter.
The contours show the continuum emission at 2.0\arcsec\ resolution and contour levels are 
$\pm$2, 4, 8, 16, 64, and 256 times the rms noise (0.0125 \mjb).  The bright star that is present here but absent in Figure \ref{fig:4526_hst} is SN1994D.
\label{fig:dust+cont}}
\end{figure}

\begin{figure}
\includegraphics[trim=0cm 0cm 1cm 8mm, width=\columnwidth, clip]{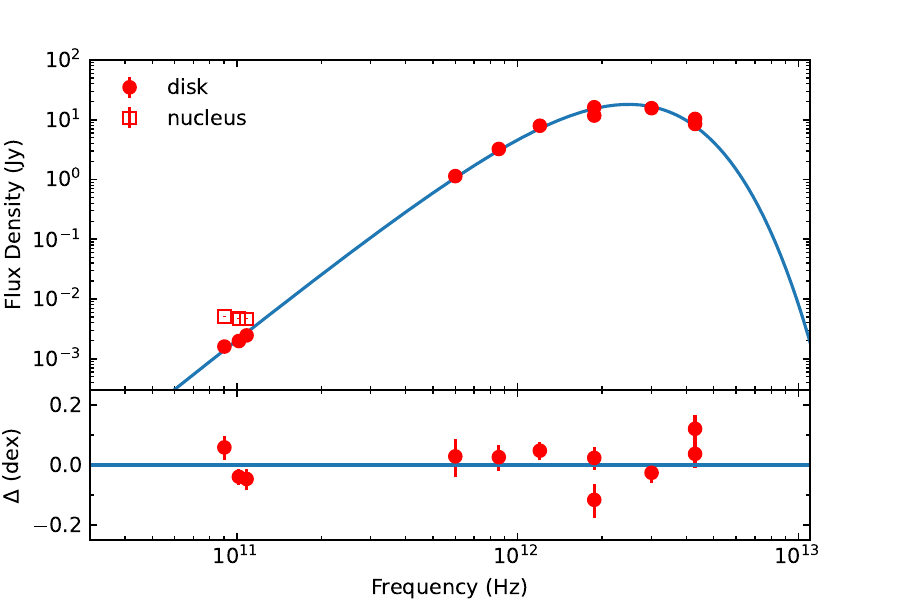}
\caption{Radio - FIR spectrum of NGC~4526.  The FIR data do not have adequate angular resolution to permit decomposition into a disk and a nucleus, so the FIR flux densities are attributed entirely to the star-forming disk.  The blue line is the modified black body fit to the FIR data from \citet{nersesian}.\label{fig:4526sed}}
\end{figure}

Imaging at a lower resolution of 2.0\arcsec\ also reveals continuum emission from a low surface brightness disk, as shown in the contours in Figure \ref{fig:dust+cont}.
The total flux density associated with NGC~4526 in the 2.0\arcsec\ image is 6.68 mJy $\pm$ 0.08 mJy (statistical) $\pm$ 0.33 mJy (absolute calibration),
of which 4.8 mJy is the point source and the remaining 1.9 mJy is the disk.  The disk emission is rather faint and uncertainties on its spectral index are large, but individual pixels have a mean spectral index of $1.8$ and an rms $\approx$ 0.7.  This strongly rising spectral index is confirmed by individual images at the extreme frequencies.  Using a high resolution image (as described above) to isolate the flux density of the point source from that of the extended dust disk, we find the dust disk to have a flux density of 1.60 $\pm$ 0.15 mJy at 90.30 GHz, 1.98 $\pm$ 0.12 mJy at 101.69 GHz, and 2.46 $\pm$ 0.20 mJy at 108.33 GHz. 
The disk and nuclear continuum flux densities are also shown in the broadband spectrum in Figure \ref{fig:4526sed}, where they can be compared to the FIR flux densities and modified black body fit 
from Dustpedia \citep{nersesian}; additional details are in Section \ref{sec:dust}.
The disk's flux density measurements are remarkably close to the long-wavelength extrapolation of the modified black body fit.

The accuracy of the extrapolation in Figure \ref{fig:4526sed} suggests that the extended 3mm disk in NGC~4526 might be, like the FIR, modified black body emission from the dust.  On the other hand, the good match in Figure \ref{fig:4526sed} could also be a coincidence, if the dust emissivity is overestimated at 3mm and a free-free contribution masks the overestimate.  Star-forming galaxies usually show a combination of free-free and dust emission at 3mm \citep{peel}.  If we interpret the disk flux densities above as a combination of flat-spectrum free-free emission and dust with an effective spectral index of 4.0, 
then we can reproduce the observed spectral index between 90 and 108 GHz as arising from a combination which is 50\% free-free and 50\% dust at 90 GHz.  Thus,
we infer that somewhere between half and all of the disk's 3mm flux density is attributable to thermal dust emission.  This fraction is somewhat larger than would be usual for starburst and spiral galaxies \citep[e.g.][]{bendo2015,bendo2016}, but as this is an early-type galaxy, the dust emission might be more strongly driven by heating from an old stellar population than by young stars.  More precise interpretations of the 3mm continuum emission in NGC~4526 will require additional data at $\approx$ 200 to 300 GHz and/or 30 GHz, to help constrain the spectral energy distribution.

\section{Gas Distribution and Kinematics}\label{sec:kin}

\begin{figure}
\includegraphics[width=\columnwidth, trim=13mm 0.8cm 1.2cm 0.7cm,clip]{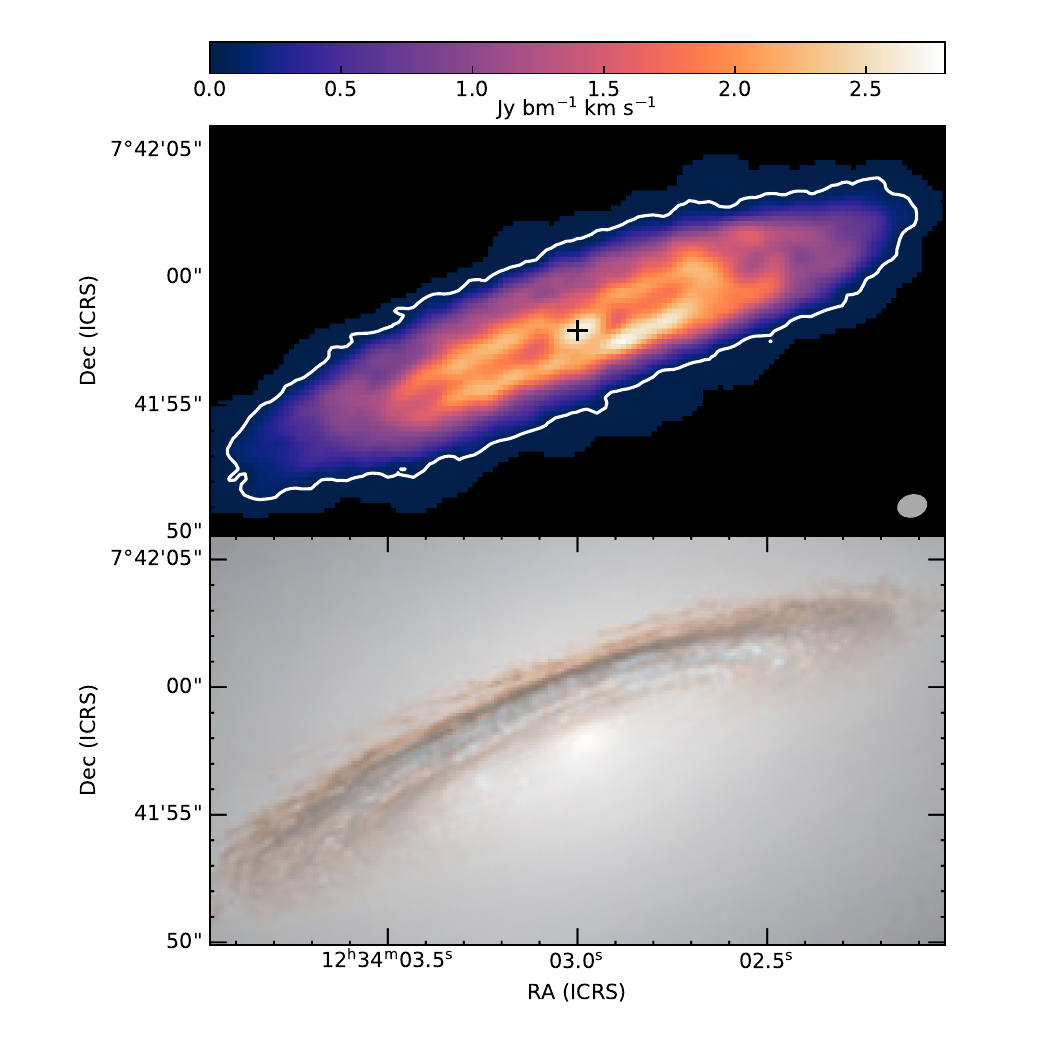}
\caption{Integrated intensity of \tweco(1$-$0) emission at 1.1$\times$0.8\arcsec\ resolution; the beam size is indicated in the lower right corner.  A cross indicates the position of the nuclear continuum source.  One contour is shown at a level of 0.05 \jybkms\ = 4.5 \solmass\persqpc\ (deprojected to a face-on surface density).  For context, the inner portion of the HST image from Figure \ref{fig:4526_hst} is reproduced at the same scale in the lower panel.
\label{fig:mom0}}
\end{figure}

Figure \ref{fig:mom0} presents an integrated \tweco(1$-$0) intensity image, produced by 
masking regions containing real emission and then integrating within the mask.
The mask is made using CASA's {\it auto-multithresh} algorithm, which clips each channel at 3$\sigma$ and then extends the clipping region spatially by about a beamwidth.
The molecular gas distribution observed here is consistent with the results shown  in \citet{davis_nature} and \citet{utomo4526}, who studied \tweco(2$-$1) at higher resolution (0\farcs3) and lower sensitivity.  We find a poorly-resolved central peak, which they showed is a fast-rotating disk or ring with a radius of about 0.5\arcsec\ (40 pc).  Outside the nuclear peak is a dip in the surface brightness, a bright molecular ring at $r=2$\arcsec\ to 7\arcsec\ (160 to 550 pc), and lower column density emission extending to about 15\arcsec\ (1.2 kpc).
The bright molecular ring is just interior to some recent star formation activity that is visible as blue stars in Figure \ref{fig:mom0} or as white regions in the unsharp-masked image in Figure \ref{fig:dust+cont}.
The peak \tweco(1$-$0) integrated intensities are 2.71 \jybkms\ in the central peak and 2.74 \jybkms\ in the molecular ring, when measured at 1.1$\times$0.8\arcsec\ resolution.  Assuming a standard conversion factor of 2.0\e{20}\persqcm\ (K \kms)$^{-1}$ or 4.3 \solmass\persqpc\ (K \kms)$^{-1}$, including He, 
and deprojecting to face-on assuming an inclination of 78\arcdeg\ (discussed below),
these values correspond to 240 \solmass\persqpc.

The peak brightness temperature in \tweco(1$-$0), at  1.1$\times$0.8\arcsec\ (75 pc) resolution, is 6.1 K; it is found in the molecular ring at a radius of 5.9\arcsec.  Somewhat lower brightness temperatures are found in the outer disk, with values of 1.0 to 3.5 K at radii of $\approx$ 12\arcsec.  Peak brightness temperatures are also quite low (1.0 to 1.5 K) in the nucleus of the galaxy, though the column densities there are high because of the large linewidths.

\begin{figure}
\includegraphics[width=\columnwidth, trim=2mm 3mm 2mm 2mm,clip]{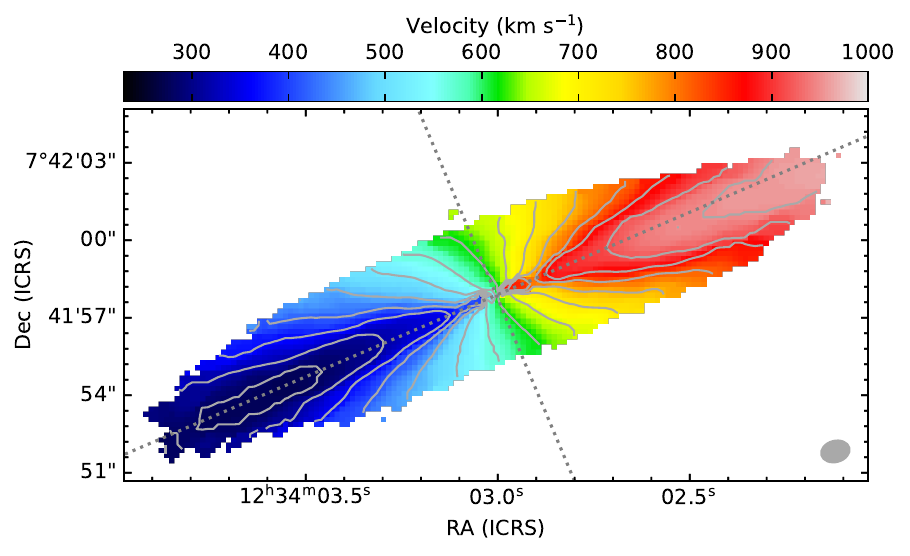}
\caption{\tweco(1$-$0) velocity field, based on the velocity of the maximum in the Gauss-Hermite profile fitted to the spectrum at each position. The kinematic major axis and its perpendicular are indicated in dashed lines.  Velocity contours are $\pm$(0, 50, 100, 150, 200, 250, 300, and 335) \kms\ with respect to systemic.\label{fig:velfield}}
\end{figure}

\begin{figure}
\includegraphics[width=\columnwidth, trim=0.3cm 1.6cm 1.4cm 1.6cm,clip]{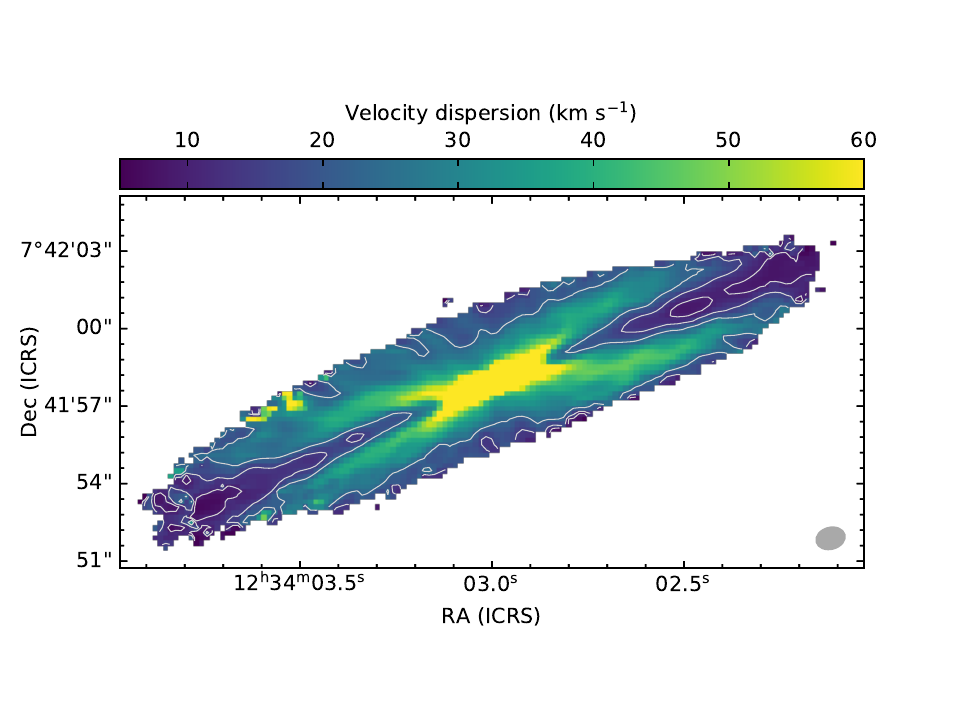}
\caption{Velocity dispersion of \tweco(1$-$0) at 1.1\arcsec\ $\times$ 0.8\arcsec\ resolution.  The dispersion is measured from a Gauss-Hermite profile fitted to the spectrum at each position.  Contour levels are 10, 15, and 22.5 \kms.   Towards the nucleus the velocity dispersions range from 250 to 300 \kms, and the kinematics there are better modeled by \citet{davis_nature}.
\label{fig:veldisp}}
\end{figure}

We created velocity fields for the molecular gas in NGC~4526 using two different methods, one using an intensity-weighted mean velocity and another by fitting 4th order Gauss-Hermite functions (i.e.\ including skew and kurtosis) to the spectrum at each location.  Significant beam-smearing effects mean that the skew and kurtosis terms are necessary to accurately reproduce the line profile shapes and identify the peak in each spectrum.
Figure \ref{fig:velfield} shows a CO velocity field and 
Figure \ref{fig:veldisp} shows the velocity dispersion.

In this paper we do not focus on detailed kinematic and dynamic analysis of the gas disk, but we still need estimates of the kinematic position angle and inclination for deprojection and analysis of radial trends in the molecular properties.  For this purpose we use a kinemetric analysis of the \tweco(1$-$0) velocity field, based on the code of  \citet{kinemetry}.\footnote{A python implementation of kinemetry is available from www.davor.krajnovic.org/software.}
The galaxy shows a well-defined kinematic center that is coincident with the 3mm continuum peak to better than 0.05\arcsec.  The kinematic position angle exhibits a slight twist from 292.2\arcdeg\ $\pm$ 0.2\arcdeg\ at $r=2$\arcsec\ to 293.8\arcdeg\ $\pm$ 0.1\arcdeg\ at $r=12.6$\arcsec.
The fitted inclination varies only from 79.0\arcdeg\ $\pm$ 0.2\arcdeg\ at $r=3$\arcsec\ to 76.3\arcdeg\ $\pm$ 0.3\arcdeg\  at $r=12.6$\arcsec.  
Thus, tilted-ring models should be highly accurate for both deprojection and dynamical analysis.

\section{Line fluxes and ratios}\label{lineratios}

\begin{figure*}
\includegraphics[width=\textwidth]{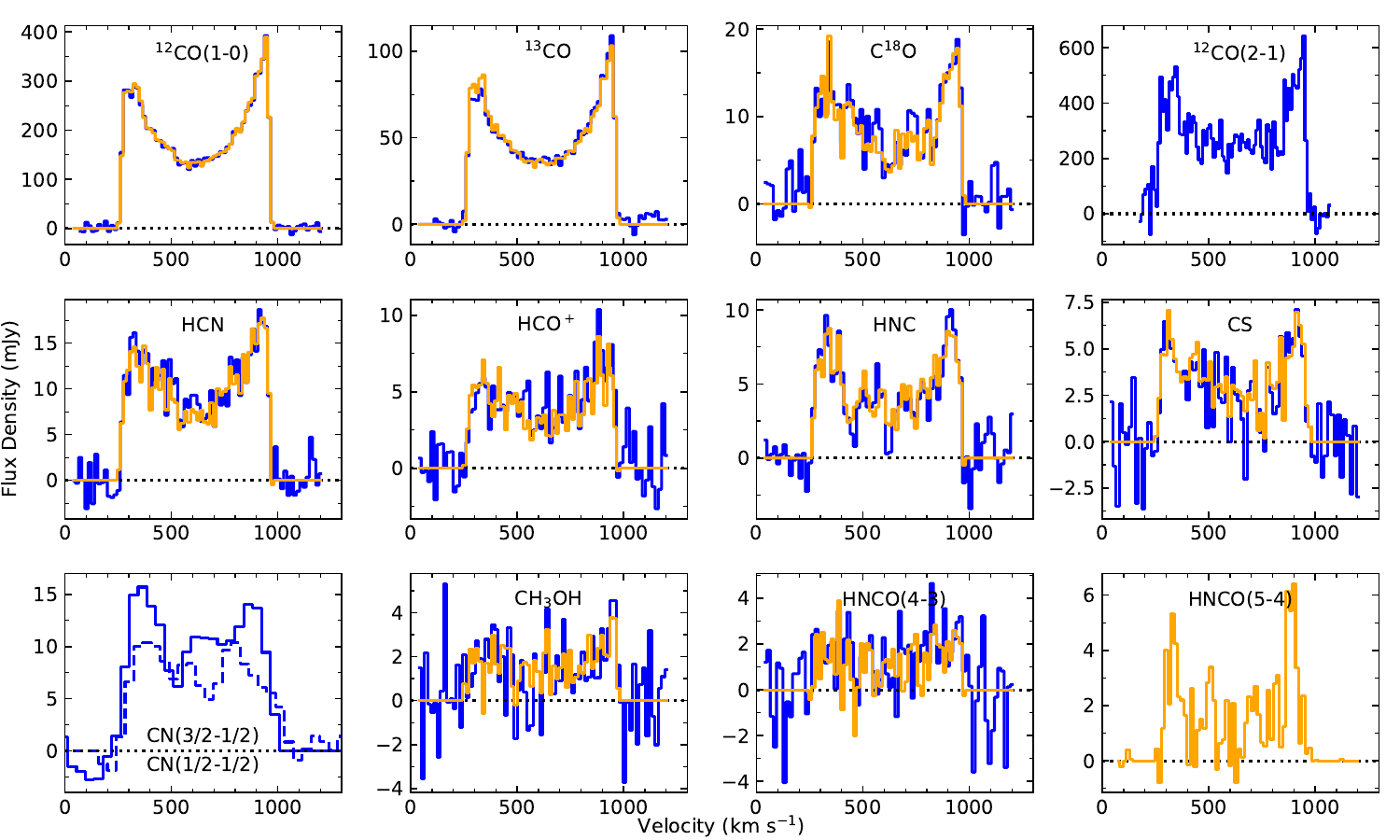}
\caption{Integrated spectra of the detected lines in NGC~4526.  The blue lines use the entire disk region at every velocity; the orange lines use a velocity-dependent mask that follows the rotation of the gas, as defined by the regions with \thirco\ emission.
The CN lines are blends of multiple components at different frequencies, and they have poorer velocity resolution, so the strict masking based on \tweco\ or \thirco\ does not work well for them.  CN(3/2$-$1/2) is shown with the solid line and CN(1/2$-$1/2) is dashed.  The HNCO(5$-$4) line partly overlaps \ceighto(1$-$0) so its spectrum is meaningless without the velocity-dependent mask.\label{fig:manyspectra}} 
\end{figure*}

\begin{deluxetable}{lllcc}
\tablecaption{Integrated line fluxes in NGC~4526\label{tab:4526lineflux}}
\tablehead{
\colhead{Line} & \colhead{Flux} & \colhead{\tweco(1$-$0)/X} \\
\colhead{}        & \colhead{\jykms} & \colhead{}
}
\startdata
\tweco(1$-$0) & 141.82 $\pm$ 0.95 & \nodata \\
\tweco(2$-$1) & 221.1 $\pm$ 5.5 & 2.56 $\pm$ 0.07 \\
\thirco(1$-$0) & 38.13 $\pm$ 0.25 & 3.40 $\pm$ 0.03 \\
\ceighto(1$-$0) & 6.50 $\pm$ 0.17 & 19.8 $\pm$ 0.5 \\
CN(N=1$-$0, J=3/2$-$1/2)     & 8.08 $\pm$ 0.31 & 17.0 $\pm$ 0.7\\
CN(N=1$-$0, J=1/2$-$1/2)     & 5.80 $\pm$ 0.13 & 23.5 $\pm$ 0.5 \\
CS(2$-$1) & 2.51 $\pm$ 0.09 & 40.8 $\pm$ 1.6 \\
\chhhoh($2_k-1_k$) &  1.21 $\pm$ 0.08 & 83 $\pm$ 6 \\    
HNC(1$-$0) & 3.46 $\pm$ 0.11 & 25.4 $\pm$ 0.8 \\
\hcop(1$-$0) & 3.00 $\pm$ 0.08 & 28.3 $\pm$ 0.8 \\
HCN(1$-$0) & 7.44 $\pm$ 0.15 & 11.3 $\pm$ 0.2 \\
HNCO$(4_{0,4}-3_{0,3})$ & 0.93 $\pm$ 0.08 & 86 $\pm$ 6 \\  
HNCO$(5_{0,5}-4_{0,4})$ & 1.34 $\pm$ 0.13 & 93 $\pm$ 9 \\
\enddata
\tablecomments{Line ratios are computed in K~\kms\ units \citep[e.g.][]{mangum+shirley}.  The uncertainties include only statistical effects, not absolute calibration.
Standard flux calibration procedures for ALMA data are usually assumed to be accurate to $\approx$ 5\% to 10\% \citep[e.g.][]{andrews2018,martin2019,almahandbook}, and ratios between lines observed at different times (see Table \ref{tab:rms}) will have these additional calibration uncertainties.
}
\end{deluxetable}

Figure \ref{fig:manyspectra} shows integrated spectra of the galaxy in all the detected lines, and Table \ref{tab:4526lineflux} gives the corresponding integrated line fluxes and line intensity ratios.
The line fluxes presented here are consistent with, though significantly higher resolution and signal-to-noise than, the corresponding values for \tweco, \thirco, \ceighto, CS, HCN, \hcop, and \chhhoh\ measured with the BIMA and IRAM 30m telescopes by Young et al (2008), Davis et al (2013) and Crocker et al (2012).
Table \ref{tab:4526ltecoldens} also gives the peak column density estimates for the various molecular species, assuming LTE at excitation temperatures of 20 K and 10 K (Section \ref{quantitative_ratios}).  Column densities are calculated using the methodology in \citet{mangum+shirley} equation 80, with molecular data from the LAMDA database.

\begin{deluxetable}{lcccc}
\tablecaption{LTE column density estimates for 2.6\arcsec\ resolution\label{tab:4526ltecoldens}}
\tablehead{
\colhead{Species}  & \colhead{Intensity}  & \colhead{log N (20 K)} & \colhead{log N (10 K)}\\
\colhead{}        & \colhead{(\jybkms)} &  \colhead{(\persqcm)} &  \colhead{(\persqcm)}
}
\startdata
\tweco &  14.59 $\pm$ 0.09 & 17.388 $\pm$ 0.003 & 17.258 $\pm$ 0.003 \\
\thirco &  3.83 $\pm$ 0.03 & 16.881 $\pm$ 0.003 & 16.746 $\pm$ 0.003 \\
\ceighto &  1.02 $\pm$ 0.03 & 16.31 $\pm$ 0.02 & 16.17 $\pm$ 0.02 \\
CN &  1.24 $\pm$ 0.05 & 13.98 $\pm$ 0.02 & 13.85 $\pm$ 0.02 \\
CS &  0.44 $\pm$ 0.03 & 13.67 $\pm$ 0.03 & 13.57 $\pm$ 0.03 \\
\chhhoh &  0.23 $\pm$ 0.03 & 13.85 $\pm$ 0.06 & 13.74 $\pm$ 0.06 \\
HNC &  0.61 $\pm$ 0.03 & 13.52 $\pm$ 0.02 & 13.37 $\pm$ 0.02 \\
\hcop &  0.52 $\pm$ 0.03 & 13.26 $\pm$ 0.02 & 13.10 $\pm$ 0.02 \\
HCN &  1.18 $\pm$ 0.03 & 13.38 $\pm$ 0.01 & 13.23 $\pm$ 0.01 \\
HNCO &  0.20 $\pm$ 0.03 & 13.90 $\pm$ 0.06 & 13.80 $\pm$ 0.06 \\

\enddata
\tablecomments{Intensities come from the peaks in the integrated line intensity images, all smoothed to match the lowest-resolution line at 2.6\arcsec\ (210 pc).  For \tweco, the intensity in column 2 refers to the J=1$-$0 transition, and for HNCO, it refers to $(4_{0,4}-3_{0,3})$.  Uncertainties reflect only statistical (signal-to-noise) effects.   All lines are assumed to be optically thin for this calculation, even though some lines are known to be thick.  The CN optical depth in the nucleus is $\tau \geq 4.5$ (Section \ref{quantitative_ratios}), and correcting for optical depth would increase the column density by at least 0.66 dex.  
HCN, HNC, and \hcop\ are probably thicker yet.}
\end{deluxetable}

\begin{figure*}
\includegraphics[width=\textwidth, trim=1cm 1cm 2cm 1cm,clip]{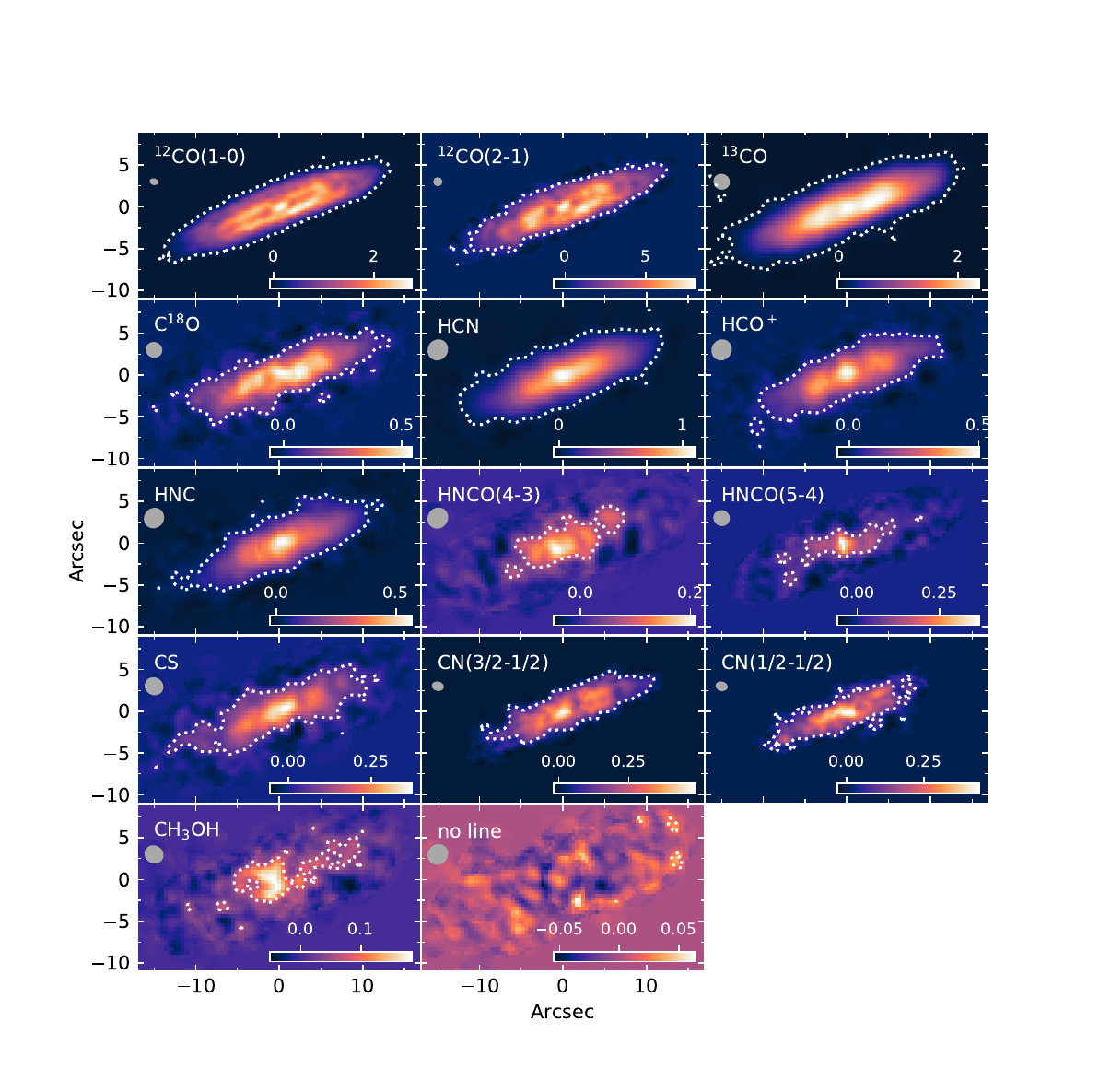}
\caption{Integrated intensities of the detected lines in NGC~4526.  These integrated intensities are computed using a velocity-dependent mask that follows the rotation of the gas, as defined by the regions with \tweco\ or \thirco\ emission.  Each line's beam is indicated by the gray ellipse in the top left corner, and the colors are scaled to the minimum and maximum of each image.  The colorbar shows integrated intensity values in \jybkms.  Dotted white contours indicate the regions where the emission is detected at a signal-to-noise ratio $> 3.$  The last panel shows the effects of our masking and integrating procedures on a line-free cube with a rest frequency of 90.0 GHz.\label{fig:4526mom0s}}
\end{figure*}

\begin{figure}
\includegraphics[width=\columnwidth, trim=4mm 3mm 1.5cm 5mm, clip]{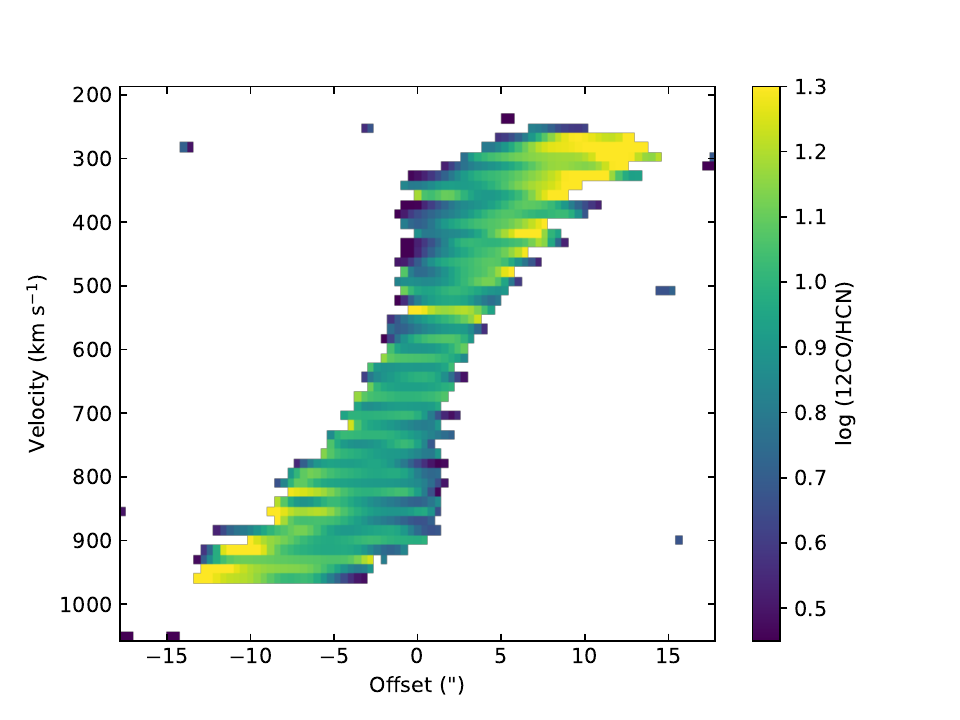}
\caption{Major axis position-velocity slice showing the \tweco(1$-$0)/HCN intensity ratio.  The \tweco\ emission is smoothed to 2.6\arcsec\ resolution, to match HCN, and the slit is 2.0\arcsec\ wide.\label{fig:pvratio}}
\end{figure}

Integrated line intensity images (Figure \ref{fig:4526mom0s}) show that the distributions of all the detected molecules in NGC~4526 are broadly similar, and the molecules differ mainly in their signal-to-noise ratio and degree of central concentration.  For example, the position-velocity slice in Figure \ref{fig:pvratio} shows that HCN is more centrally concentrated than \tweco, as the \tweco/HCN ratio increases with radius.

\begin{figure}
\includegraphics[width=\columnwidth, trim=8mm 4mm 14mm 6mm,clip]{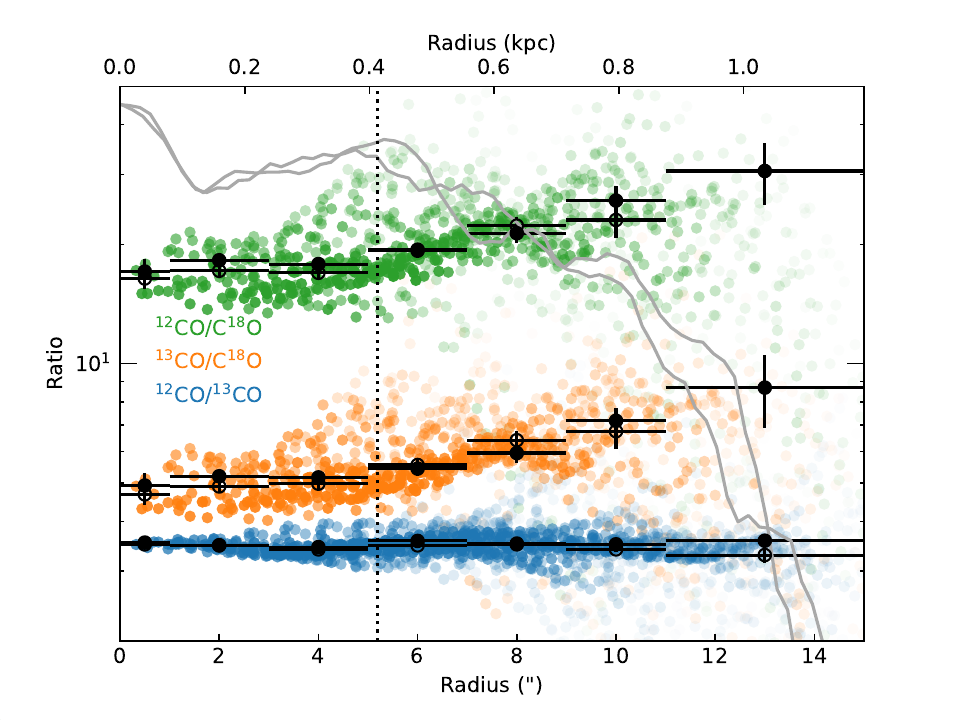}
\caption{Colored circles indicate line ratios computed at each pixel from the integrated intensity images (method 1 of section \ref{lineratios}); fainter symbols have lower signal-to-noise ratio.  In this figure, \tweco\ is always (1$-$0).  Pixels are plotted at their deprojected radius.  Black filled symbols with error bars correspond to ratios computed from the position-velocity slices (method 2) and black open symbols with error bars show ratios computed from the annular spectra (method 3).  Grey lines indicate the surface brightness profile of the molecular gas, based on a folded major axis slice of the high resolution \tweco(1$-$0) integrated intensity image, arbitrarily scaled for visibility.  The dotted vertical line marks the approximate location of the local surface brightness peak in the CO ring.
\label{scatterplot}}
\end{figure}

\begin{figure}
\includegraphics[width=\columnwidth, trim=8mm 4mm 14mm 6mm,clip]{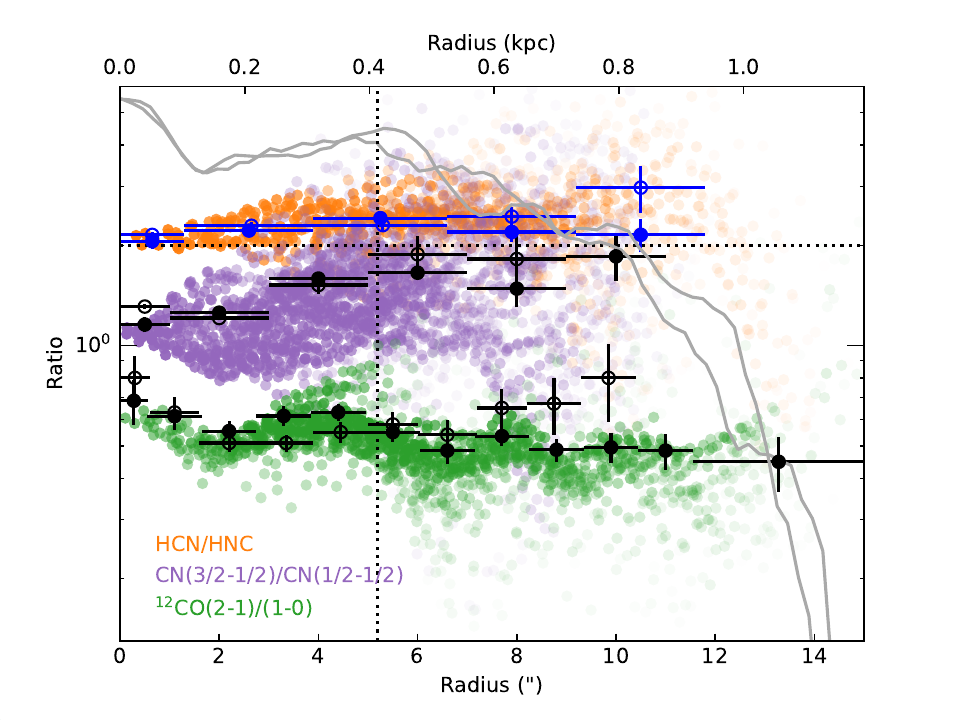}
\caption{Similar to Figure \ref{scatterplot}, for additional selected ratios.
The dashed horizontal line indicates the theoretical optically thin LTE ratio of the two CN blends.\label{scatterplot2}}
\end{figure}

\begin{figure}
\includegraphics[width=\columnwidth, trim=6mm 4mm 1.5cm 3mm, clip]{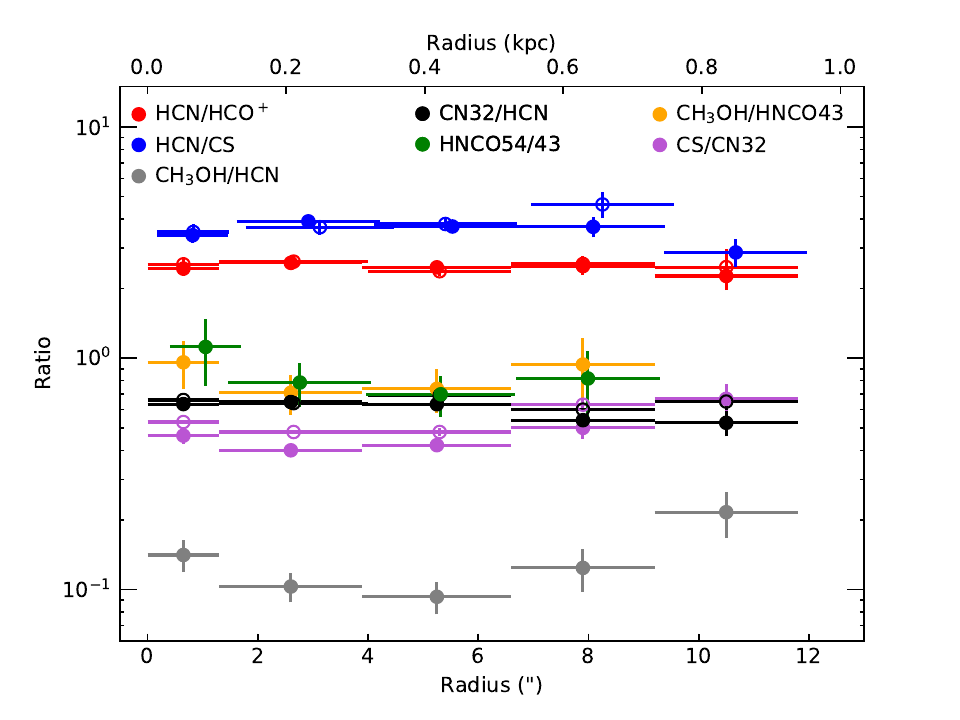}
\caption{Radial variations in some line ratios in NGC~4526.  As in Figure \ref{scatterplot}, filled symbols are ratios computed from the position-velocity slices and open symbols are from the annular spectra.  Small radial offsets are added to some points for clarity.  `CN32' refers to the (J=3/2$-$1/2) blend. \label{radialratios1}}
\end{figure}

\begin{figure}
\includegraphics[width=\columnwidth, trim=6mm 3mm 1.5cm 3mm, clip]{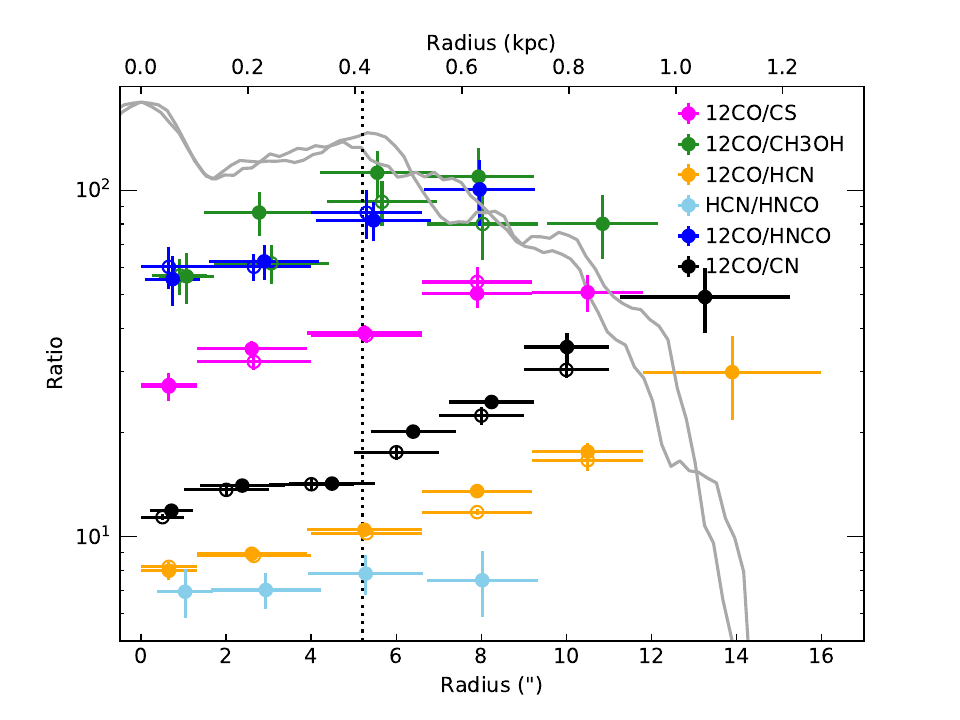}
\caption{Radial variations in more molecular ratios.  Lines and symbol types have the same meanings as in Figure \ref{scatterplot} and \ref{radialratios1}; ratios are grouped into these figures purely for purposes of dynamic range and legibility.  In this figure, \tweco\ is always the (1$-$0) transition.
\tweco/CN here uses only the CN(J=3/2$-$1/2) blend, and HNCO uses the $(4_{0,4}-3_{0,3})$ line.
\label{radialratios2}}
\end{figure}

We quantify the radial variations in the line ratios using three main techniques which all give similar answers.
In method 1 we create matched-resolution integrated intensity images, compute a line ratio directly for every pixel, and use the orientation and inclination of the disk to associate each pixel with its deprojected radial distance.
Method 2 is to construct matched-resolution major axis position-velocity slices (e.g.\ Figure \ref{fig:pvratio}), integrate over radial regions, and compute the ratios of the corresponding sums in radial bins.  Uncertainties for this method can be estimated from standard error propagation on the number of independent data points in the bin and also by varying the clip threshhold used to select pixels with signal.  (We always use the same data volumes for both lines, to eliminate biases related to different line strengths.)  
Method 3 is to construct integrated spectra by summing the data cube over elliptical annuli of the appropriate position angle and inclination.  We then carry out a least-squares optimization modeling a fainter line's annular spectrum as a multiple of the brighter line.  This technique works because the distributions of all the species are broadly similar, so that the corresponding annular spectra all have the same shape.  The uncertainty in the ratio can be characterized with standard $\chi^2$ goodness-of-fit criteria.  
Figures \ref{scatterplot}  and \ref{scatterplot2} show comparisons of these methods for the CO isotopologue ratios and for a few other selected line ratios, indicating good agreement for all of them.
Figures \ref{radialratios1} and \ref{radialratios2} show radial trends in other line ratios.

In this section we present a large-scale overview of the relative intensities of the molecular lines in NGC~4526 with some qualitative comments on unusual aspects.  In Section \ref{quantitative_ratios} we undertake more quantitative analysis of the physical conditions in the molecular gas.  To summarize the qualitative aspects, we note that
NGC~4526 is relatively bright in \thirco\ and the high-density tracers (HCN, \hcop, HNC, CN, and CS), when compared to many spirals.  It has relatively faint emission in the shock tracers \chhhoh\ and HNCO, though
it should be noted that existing observations of those species are biased towards active and starbursting galaxies.
All of the line ratios between \tweco\ and other molecules, with the singular exception of \thirco, show strong radial gradients of factors of two to four over 1 kpc in radius.
In contrast, 
there is limited or unmeasurable radial variation in \tweco(1$-$0)/\thirco(1$-$0) and the line ratios of any two high density tracers, such as HCN/\hcop, HCN/HNC, HCN/CS, \hcop/CS, HCN/HNCO, and \chhhoh/HNCO.

\subsection{CO Isotopologues}

\begin{figure}
\includegraphics[width=\columnwidth, trim=5mm 3mm 1cm 5mm, clip]{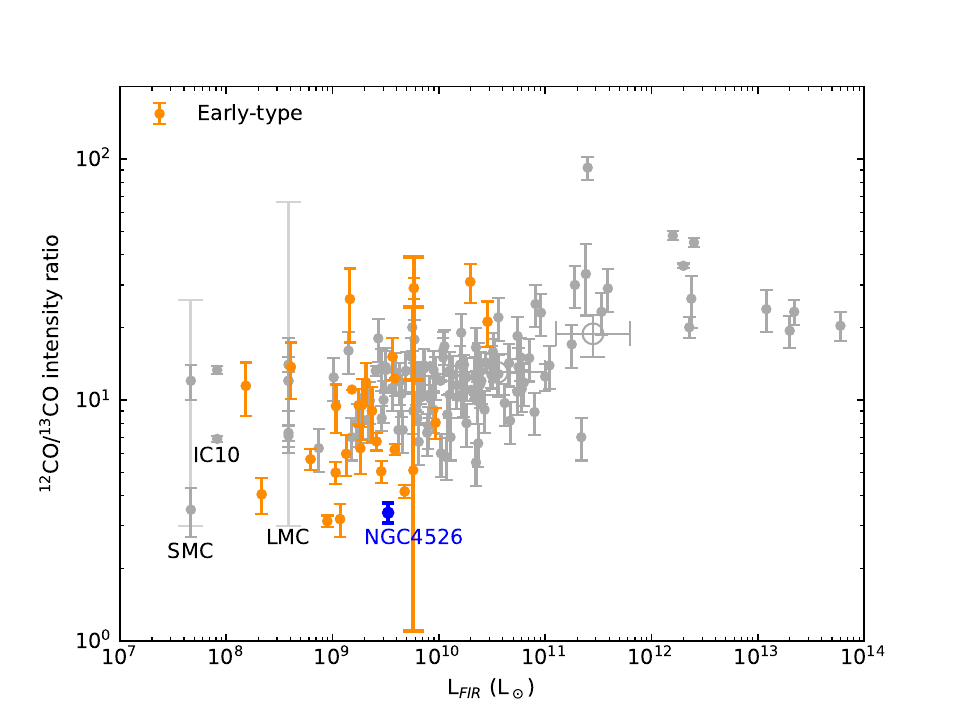}
\caption{A compilation of observed \tweco/\thirco\ line ratios, in the same spirit as Figure 1 of \citet{zhang2018} and \citet{me_7465}.  NGC~4526 is highlighted in blue and other early-type galaxies are orange; other galaxy types are grey.
Data are drawn from \citet{davis_13co}, \citet{jd_13co}, \citet{brown+wilson}, \citet{cormier18}, \citet{henkel14}, \citet{braine2017}, \citet{heikkila99}, \citet{zhang2018}, \citet{israel03}, \citet{mendez-hernandez},
\citet{crocker_hd}, \citet{a3d_13co}, and \citet{israel2020}.  For early-type galaxies, the largest error bar belongs to Cen~A \citep{mccoy_cena} and it shows a range in measured values rather than an uncertainty.
\label{fig:lir2}}
\end{figure}

The \tweco(1$-$0)/\thirco(1$-$0) ratio observed in NGC 4526 is $3.4\pm0.3,$ which is one of the lowest values observed outside of the Local Group (Figure \ref{fig:lir2}).
Beyond the Milky Way, such low values are rare and noteworthy; some examples include NGC~4429 and NGC~4459 (two other Virgo Cluster early-type galaxies) and parts of resolved galaxies like the SMC and LMC, Centaurus A, and IRAS 04296+2923 \citep[citations in Figure \ref{fig:lir2} and][]{meier2014}.
Spatial variations in \tweco/\thirco\ in NGC~4526 are also remarkably small, being less than 5\% and smaller than the corresponding statistical uncertainties.
Resolved studies of \tweco/\thirco\ within spiral galaxies usually show greater variation than that, sometimes increasing with radius and sometimes decreasing \citep{carmasting,cormier18,topal_ngc628}, though usually over larger radial ranges.

\begin{figure}
\includegraphics[width=\columnwidth, trim=5mm 3mm 1cm 5mm, clip]{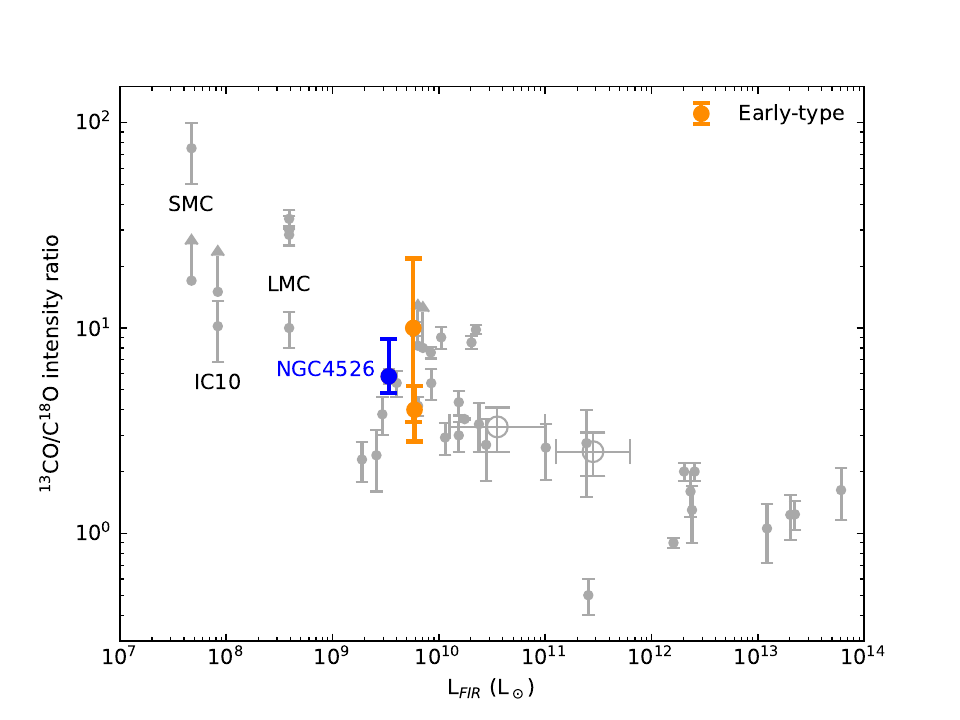}
\caption{A compilation of observed \thirco/\ceighto\ line ratios, after Figure 1 of \citet{zhang2018}.  Data are also drawn from \citet{johansson94}, \citet{wang09}, \citet{denbrok22}, and works noted in Figure \ref{fig:lir2}.  For NGC~4526, the error bars represent the range of values observed in our data (since that range is larger than the uncertainties).
\label{fig:lir}}
\end{figure}

The \thirco/\ceighto\ line ratios in NGC~4526 are not particularly unusual, as they 
range from 4.7$\pm$0.3 to 8.7$\pm$1.8, and they
fall near the median of the values observed in other nearby galaxies (Figure \ref{fig:lir}).
But unlike \tweco/\thirco, the \tweco/\ceighto\ and \thirco/\ceighto\ line ratios have significant radial structure. 
Figure \ref{scatterplot} shows that the ratios involving \ceighto\ rise by a factor of 1.8 over this compact disk.  The figure also identifies the location of the bright molecular ring, and the gradients in the \ceighto\ ratios are markedly larger exterior to the ring than they are interior to the ring.  Thus the rise from \thirco/\ceighto\ = 5 to 9 occurs over the radial range of 0.4 to 1.0 kpc.  For comparison, the same rise in the Milky Way \thirco/\ceighto(1$-$0) line ratios occurs over a radial range of about 6 to 12 kpc \citep{wouterloot}.  Additional comparisons to other spiral galaxies are discussed in more detail in Section \ref{sec:disc_radial}.

Resolved measurements of the \tweco(2$-$1)/(1$-$0) line ratio mostly range between 0.4 and 0.7 (Figure \ref{scatterplot2}).
These values are very typical for spiral galaxies \citep{leroy_jlevels,vertico}.  Indeed, the star formation efficiency of the molecular gas in NGC~4526 is consistent with values in many of those spiral galaxies even though its gas content is lower than most of them.
Evidence for any spatial variations in the (2$-$1)/(1$-$0) line ratio is not compelling; the ratio might be slightly elevated towards the nucleus, but because of the uncertainty associated with continuum subtraction in the (2$-$1) data (Appendix \ref{carmadata}), the issue should be re-evaluated with new data.
Like NGC~4526, many spirals show small radial variations in the (2$-$1)/(1$-$0) line ratio, and sometimes the spirals show arm/interarm variations \citep{denbrok21,denbrok22}.

\subsection{High Density and PDR tracers}

The range of \tweco/HCN values we measure in NGC~4526 is 8.0$\pm$0.5 to 30$\pm$8. 
The values in the outer disk of NGC~4526 are comparable to those in the 
spirals in the EMPIRE survey \citep{jd_bigsurvey},  but only a handful of points in the spirals have bright enough HCN to match the center of NGC~4526.  Of course, it is worth noting that the EMPIRE data have lower resolution that we are using here (1 to 2 kpc compared to 200 pc).   Similar results also hold for CN and CS.
\tweco/CS in NGC~4526 ranges from 27$\pm$2 to 51$\pm$6; 
the spirals in \citet{gallagher2018} have typical values of \tweco/CS $\sim$ 90 in their central kpc.
For \tweco/CN ratios (isolating just the J=3/2$-$1/2 blend), we measure values ranging from 11.4$\pm$0.2 to 49$\pm$10, or 17$\pm$0.7 for a global line ratio.  Other global \tweco/CN ratios measured by \citet{wilson_cn} range from 20 to 150. 
In short, the high density tracers in NGC~4526 are not exceptionally bright in comparison to \tweco\ but, especially for CS, they are factors of  roughly 2 brighter than the median values for nearby spirals.

The CN radical is expected to be formed through photodissociation of HCN and/or through reactions involving C and C$^+$, so it should be enhanced in regions with strong radiation fields.  It may also be particularly enhanced in XDRs \citep{boger_cn, meijerink2007}.
\citet{ledger_cn} find some supporting evidence for factors of two enhancements in CN/HCN ratios towards the nuclei of some starbursting galaxies and ULIRGs. 
In contrast, we find no such evidence for variations in CN/HCN in NGC~4526, or in CN/CS or HCN/\hcop\ for that matter, even though there is an AGN in the center of the galaxy and a faint X-ray point source as well \citep{chandra_atlas}.  The AGN in NGC~4526 may simply be faint enough that its effects on molecular line ratios are restricted to regions smaller than our current resolution.

\begin{figure}
\includegraphics[width=\columnwidth, trim=0.3cm 1.7cm 0cm 3cm, clip]{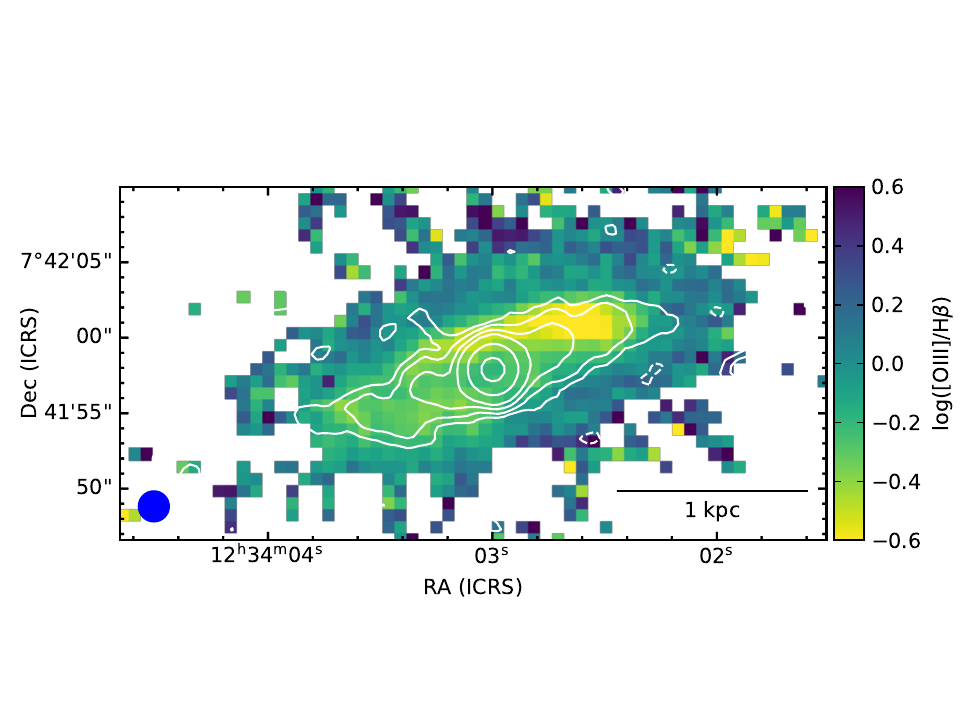}
\caption{3mm continuum emission contours, overlaid on the \oiiihb\ line ratio from the SAURON/\atlas\ data \citep{sarzi06}.  
Contours are as in Figure \ref{fig:dust+cont}.  The prominent \oiiihb\ minimum is coincident with young blue stars but not with any enhanced continuum; these observations may support our interpretation that the 3mm continuum is mostly dust rather than free-free emission.
\label{fig:oiiihb}}
\end{figure}

In the light of evidence that CS emission is enhanced in PDRs \citep[e.g.][]{lintott_cs,meier2005}, 
it has been suggested that CS might trace star formation activity even better than a standard dense gas tracer such as HCN.  For example, \citet{davis_cs} found that CS is enhanced relative to HCN in galaxies whose ionization is more strongly dominated by star formation activity than by AGN.  That study used unresolved single-dish measurements of the molecular lines and traced star formation activity through integrated \oiiihb\ line ratios.
With our new ALMA data we can also test whether this trend applies in an internal, spatially-resolved manner.  NGC~4526 displays a wide range of \oiiihb\ values (Figure \ref{fig:oiiihb}), from log(\oiiihb) = $-0.85$ in its northwest quadrant to $-0.2$ in its nucleus and even higher values outside the molecular disk.
However, we find no evidence of variable CS/HCN ratios anywhere in the galaxy, and no significant difference between the northwest quadrant and other parts of the galaxy.  While the integrated measurements in NGC~4526 are consistent with a trend between low \oiiihb\ and high CS/HCN, the internally resolved measurements do not follow such a trend.  

The above discussion also highlights the fact that it is not yet clear whether internal variations of \oiiihb\ in NGC~4526 trace stochasticity in the star formation rate in the last $\sim$ 10 Myr, shocks, metallicity variations, or something else.
The lowest \oiiihb\ ratios are on the receding side of the galaxy, which is the side with brighter CO emission (Figure \ref{fig:manyspectra}).  However, the asymmetry in \oiiihb\ is not reflected in the 3mm continuum emission or in the 24\micron\ dust emission \citep{spitzerwbendo}. Additional work on other star formation tracers will be required before NGC~4526 can give a clear conclusion about CS/HCN and star formation activity or about internal variations in the star formation efficiency.

\subsection{\chhhoh\ and HNCO -- shock tracers}
 
Our new measurements of \tweco/\chhhoh\  in NGC~4526 broadly confirm the single-dish measurements of \citet{davis_cs}.  NGC~4526 is relatively faint in \chhhoh\ emission compared to the other early-type galaxies in that sample.  We find hints that the distribution of \chhhoh\ in NGC~4526 may be different from that of the other molecules; the integrated intensity (Figure \ref{fig:4526mom0s}) suggests a broader extent of \chhhoh\ in the minor axis direction than that of the other species, so the \tweco/\chhhoh\ ratios in the galaxy are systematically higher on the major axis ($\sim$ 110)  than on the minor axis ($\sim$ 40).
But \chhhoh\ is very faint and this suggestion should be verified with higher sensitivity data.  Similarly, there are hints that some tracers such as \chhhoh\ and HNCO(5$-$4) are brighter in the far-receding and near-approaching quadrants of the galaxy than in the other two quadrants, suggestive of a bar structure, and those features will be explored in a future paper.

\citet{meier2005} have argued that HNCO is also a shock tracer, like \chhhoh.
At present there are few other HNCO detections in early-type galaxies, but for comparison purposes we note that
the \tweco/HNCO(4$-$3) and HCN/HNCO(4$-$3) line ratios in NGC~4526 have similar values to those found in the nearby starburst galaxy NGC~253 \citep{meier2015} and in the early-types NGC~4710 and NGC~5866 \citep{topal}.
Interestingly, NGC~253 also suggests that the low-J transitions of HNCO might be anticorrelated with vigorous star formation activity; 
\citet{meier2015} found a central deficit in HNCO(4$-$3), which they interpreted as a signature of excitation effects in high-temperature molecular gas and/or photodissociation of HNCO in the vicinity of the starburst.
In NGC~4526, in contrast, we find {\it higher} HNCO(4$-$3) intensities (relative to \tweco) in the center of the galaxy, but constant HCN/HNCO(4$-$3) ratios.  
The interpretation of this radial trend is not entirely clear.  If the majority of the star formation activity is just exterior to the molecular ring, as suggested by the young blue stars and the \oiiihb\ ratios, then the lower HNCO intensities in the outer disk of NGC~4526 could be consistent with high-temperature and/or photodissociation effects in PDRs.  But the constancy of HCN/HNCO(4$-$3) instead suggests that PDR effects are not the dominant drivers of the radial behavior of HNCO in NGC~4526.

\section{Physical conditions in the molecular gas}\label{quantitative_ratios}

\subsection{CO isotopologues}\label{sec:coradexresults}

We have spatially resolved data on four CO isotopologue transitions [ \tweco(1$-$0), \tweco(2$-$1), \thirco(1$-$0), and \ceighto(1$-$0) ] plus a single-dish measurement of the \tweco(2$-$1)/\thirco(2$-$1) line ratio 
\citep[4.6$\pm$0.7;][]{crocker_hd}, and
we use these data to explore the physical conditions in the CO-emitting gas.
We use the RADEX code \citep{radex} to compute predicted line ratios for a variety of physical conditions and a Markov chain Monte Carlo sampler \citep{emcee} to estimate maximum likelihood parameters with their uncertainties.  The details are in Appendix \ref{radex}, and we summarize the results here. 

The conditions in the CO-emitting gas of NGC~4526 are most tightly constrained in the molecular ring.
There we estimate the column density of \tweco\ to be log N(\tweco) [cm$^{-2}$] = $18.0 \pm 0.3.$ (Strictly speaking, RADEX constrains the ratio of the column density to the linewidth, $N/\Delta v.$   We assume a linewidth of 30 \kms\ in this part of the galaxy, based on the velocity dispersions along the major axis in Figure \ref{fig:veldisp}.)  The quoted uncertainty contains 68\% of our inferred probability distribution.
We also find the \htoo\ volume density to be log \nhtoo[cm$^{-3}$] = $2.4 \pm 0.4$ and the kinetic temperature to be \tkin\ $> 10$ K.  The density and temperature are
still not particularly well constrained since we do not have observations of the (2$-$1) transitions in an optically thin line.  Despite that limitation, the isotopologue abundances can be constrained.
We find the [\tweco/\thirco] abundance ratio to be $7.8^{+2.7}_{-1.5},$ an unusually low value; values between 20 and 60 are much more common for nearby spirals (Section \ref{sec:disc_global}).  We find the optical depth of \thirco(1$-$0) to be a relatively high value of $\tau_{13} = 1.1^{+0.9}_{-0.5}$.  
The derived value of the [\thirco/\ceighto] abundance ratio in the molecular ring is $6.2^{+0.8}_{-0.6}$.

We also find strong evidence for a radial gradient in [\thirco/\ceighto], as we 
we infer values of 5.7$^{+0.7}_{-0.6}$ in the nucleus ($r \lesssim 2$\arcsec) 
 and 9.2$^{+1.0}_{-1.0}$ at the outer edge of the disk. 
Those abundances are approximately 15\% larger than the
corresponding observed line intensity ratios because of the optical depths in \thirco(1$-$0).

We find no particularly strong evidence for a gradient in [\tweco/\thirco].
Our inferred value for the outer disk is $6.5^{+3.0}_{-1.3},$ which is entirely consistent with the measurement in the ring.  This abundance ratio is more poorly constrained in the nucleus because of the lower brightness temperatures and the broader allowed range of kinetic temperatures, but the marginalized probability distribution peaks at [\tweco/\thirco] $\sim 6.3$ in the nucleus.
The lack of a gradient in [\tweco/\thirco] is a bit surprising, as there is some expectation for lower [\twec/\thirc] ratios in the centers of galaxies.
Measurements and models for the Milky Way are presented by \citet{romano2019}, for example.  Further discussion of abundance ratio gradients is in Section \ref{sec:disc}.

As is commonly found in other galaxies, the conditions that reproduce the CO isotopologue intensities in NGC~4526 do not reproduce the observed intensities of the high-density tracers.
For example, using the density, temperature, and column density appropriate to the molecular ring, and 
typical abundances of [CO/\htoo] $\approx 10^{-4}$ and [HCN/\htoo] $\approx 10^{-8}$ \citep{meier2015}, we  predict HCN(1$-$0) intensities on the order of 1 mK.  The observed intensities are roughly 0.5 K.  Thus, these data are consistent with the usual paradigm that the high-density tracers arise in a significantly different ``phase" of molecular gas from that traced by the CO isotopologues.

\subsection{HNCO}\label{sec:hnco_temp}

We detect two transitions of HNCO,  $(4_{0,4}-3_{0,3})$ and  $(5_{0,5}-4_{0,4})$, whose rest frequencies are 87.925237 GHz and 109.9057490 GHz respectively.  They provide some constraints on the excitation temperature of the high-density molecular gas.
The (5$-$4) transition requires careful treatment as it is close to \ceighto(1$-$0), and the velocity width of NGC~4526 necessitates careful masking.

The  HNCO (5$-$4)/(4$-$3) line ratio measured from the global spectra (Figure \ref{fig:manyspectra}) is 0.92 $\pm$ 0.11.  There is no evidence for significant radial variation (Figure \ref{radialratios2}).
In optically thin LTE, that ratio implies excitation temperatures in the range of 7.4 K to 12.5 K \citep[][Eq 27 and 32]{mangum+shirley}.  Simple estimates of the HNCO optical depths suggest that they are probably around 1, as we discuss in Section \ref{sec:tau_cn}; if this assumption is accurate, the allowed temperatures fall in a slightly expanded range from 5.4 K to 13.4 K.
Of course, higher temperatures are required if the density is too low to establish LTE.

This excitation temperature estimate from HNCO is consistent with the conditions inferred from CO above, which were \tkin\ $> 10$ K.  
\citet{ryen_spire} also found an excitation temperature of 15.7 $\pm$ 2.0 K from the [\ion{C}{1}] (1$-$0) and (2$-$1) lines in NGC~4526, and the dust temperature from the far-IR spectral energy distribution in Figure \ref{fig:4526sed} is 25.1 $\pm$ 0.8 K \citep{nersesian}.  As noted by \citet{hacar_hnc}, we might expect somewhat higher temperatures inferred from CO or [\ion{C}{1}] than from a high-density tracer like HNCO, as CO and [\ion{C}{1}] should be preferentially emitted from the lower extinction, warmer outer regions of clouds.  Our temperature estimates in NGC~4526 confirm this general picture.

\subsection{HCN/HNC}

HCN/HNC line ratios in NGC~4526 are consistent with a constant value everywhere in the disk; of the pixels with detections in both lines, the median ratio is 2.4 and the dispersion is 0.5.  Binned measurements range from 2.06 $\pm$ 0.02 to 3.2 $\pm$ 0.6 at $r < 12$\arcsec.  These values are not unusual for extragalactic sources \citep[e.g.][and references therein]{eibensteiner}. 
The critical densities of the two species are similar, so we do not expect substantial differences in their excitations.  

Several authors \citep[e.g.][]{hirota_hnc,graninger_hnc,hacar_hnc} have argued that the primary driver of variations in HCN/HNC line ratios, for typical interstellar molecular conditions,
will be abundance variations caused by preferential destruction of HNC at warmer temperatures.  At the low temperatures relevant here, the mechanism should be gas-phase reactions of HNC with neutral atoms.  Hacar et al.\  demonstrate that HCN/HNC line ratios are an effective indicator of the gas kinetic temperature, and they provide a calibration relating the observed line ratio to the temperature.  For the line ratios measured in NGC~4526 the inferred temperatures are 20 to 24 K over most of the disk.

The calibration in Hacar et al.\ is derived from studies in Orion, and it is also roughly consistent with results from other molecular clouds \citep{jin_hnc} and protoplanetary disks \citep{graninger_hnc2}.  
However, there is some doubt about whether extragalactic observations, which naturally include a wide variety of environments, can also be interpreted using a calibration derived from individual molecular clouds.
The temperatures inferred from the HCN/HNC thermometer in NGC~4526 are roughly a factor of 2 larger than those inferred from the HNCO transitions.  
Discrepancies of similar magnitude are found in other extragalactic studies \citep{eibensteiner}, who also conclude that HCN/HNC may not be reliable for extragalactic temperature measurements.
In principle, the HNCO measurement is less susceptible to systematic uncertainties as it utilizes two transitions from the same molecule.  On the other hand, in our observational setup, the HNCO measurement is more susceptible to calibration uncertainties as the lines are fainter and they were not observed simultaneously like HNC and HCN were.  But all the available estimates suggest relatively low temperatures for the molecular gas in NGC~4526, with little to no evidence for radial trends.

\subsection{CN hyperfine blends and implications for the other high-density tracers}\label{sec:tau_cn}

The 3mm ($N=1-0$) spectral lines of CN consist of a set of hyperfine components which are grouped into two main blends, one with \jthreetwo\  around 113.49 GHz and one with \jonetwo\ around 113.19 GHz.  
The ratios of these two blends can be used to infer the optical depths of the CN lines; the optically thin limit in LTE is 
I(\jthreetwo)/I(\jonetwo) = 2.0, so deviations from that value imply significant optical depths \citep[e.g.][]{tang2019}.
Previous observations suggested that most observed $^{12}$CN is optically thin \citep[][and references therein]{wilson_cn},
though \citet{alchemi1} find evidence for optically thick CN emission in NGC~253, and \citet{tang2019} found optical depths in the \jthreetwo\ component as high as $\tau_{3/2} = 1.9$ in their sample of three starburst galaxies.

Here we find even lower line ratios and higher optical depths in the center of NGC~4526.
Figure \ref{scatterplot2} shows values for this CN blend ratio as low as 1.12 $\pm$ 0.16 in individual pixels in the center of the galaxy, or 1.16 $\pm$ 0.06 computed from the position-velocity slices, rising to values consistent with 2.0 (the optically thin limit) at $r > 5$\arcsec.

In the LTE assumption, the low CN hyperfine blend ratios require $\tau_{3/2} \ge 4.5$ in the nucleus of NGC~4526. 
 For typical abundances and column densities, the CN ratio also suggests an excitation temperature $\leq 20$ K in the nucleus of NGC~4526; higher temperatures depopulate those low energy levels and the transitions become optically thin.  This low temperature is also
 roughly consistent with the temperatures inferred from CO, [\ion{C}{1}] and HNCO.  
If the excitation temperature is constant with radius, as suggested by the HNCO(5$-$4)/(4$-$3) line ratio, then the radial decrease in CN optical depth requires radial decreases in \nhtoo\ of a factor of 30, or in N(CN) of a factor of 100, or some combination of both.  If we also take the HCN/\hcop\ ratio to be an indicator of density, then its constancy with radius suggests the CN optical depth changes should be attributable mostly to the CN column density, and perhaps to variations in the CN abundance.

Large optical depths in CN in the center of NGC~4526 also imply large optical depths in other high-density tracers.  For example, if the abundance ratio [CN/HCN] falls in the range of 1 to 5 \citep[e.g.][]{liszt+lucas,meier2015}, and if LTE applies, then we expect the optical depth of HCN(1$-$0) to be between 1.2 and 6 times that of the CN 113.49 GHz blend.  In other words, we expect the optical depth of HCN(1$-$0) to be at least in the range of 5 to 30.  Similar comments apply to \hcop\ and HNC.  \citet{jd_hcn} found comparably high optical depths in HCN and \hcop\ in some nearby spirals, based on observations of those species' \thirc\ isotopologues.

High optical depths might explain the unusual lack of radial trends in any of the ratios involving two high-density tracers in NGC~4526.  Implications for the use of an HCN/HNC thermometer 
such as that in \citet{hacar_hnc} are not clear, since that proposed thermometer is based on a purely empirical calibration.  In contrast, if [CN/HNCO] is $\sim$ 1 \citep{meier2015}, the optical depth of the HNCO(5$-$4) and (4$-$3) transitions should be between 0.2 and 0.4 times that of CN.  Thus, the HNCO lines are probably somewhat thick ($\tau \sim 1$) in the nucleus of NGC~4526 and the HNCO temperature estimate from section \ref{sec:hnco_temp} should still be reliable.

\section{Dust and CO/\htoo\ conversion factors}\label{sec:dust}

With an assumed dust emissivity and a gas-to-dust mass ratio, the total dust and \htoo\ masses can be obtained from dust continuum emission \citep[e.g.][and many others]{hildebrand83,leroy_alphas}.
In the case of NGC~4526, we have argued (Section \ref{sec:cont}) that at least half of the 3mm continuum is dust emission, so the image in Figure \ref{fig:dust+cont} can give loose constraints on the resolved \htoo\ column density estimates and CO/\htoo\ conversion factors.
Here we assume all of the extended 3mm continuum is dust emission.
We use the modified black body emissivity model for NGC~4526 from the Dustpedia results \citep{nersesian}:
they find a dust temperature of 25.1$\pm$0.8 K and a mass of (4.38$\pm$0.58)\e{6} \solmass, assuming a distance of 15.35 Mpc.  These values derive from an emissivity of 0.640 m$^2$ kg$^{-1}$ $(250 \mu m/\lambda)^{1.79}$, which also predicts 7.4\e{-3} m$^2$~kg$^{-1}$ at 99 GHz and gives a remarkably accurate prediction of the observed 3mm flux density.  We also assume a \htoo/dust mass ratio of 100, based on the results of \citet{cortese2016} for Virgo Cluster and/or \hi-deficient galaxies of stellar mass above $10^{10.5}$ \solmass.

\begin{figure}
\includegraphics[width=\columnwidth, trim=5mm 3mm 1cm 5mm, clip]{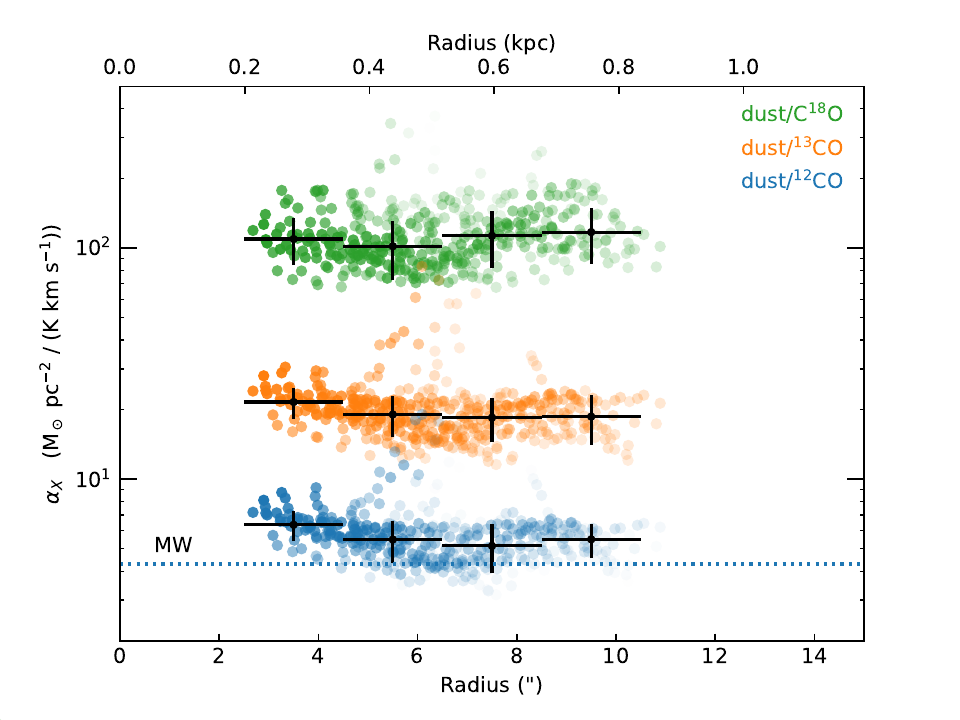}
\caption{Colored points are individual pixels' \tweco(1$-$0)/\htoo, \thirco(1$-$0)/\htoo, and \ceighto(1$-$0)/\htoo\ conversion factor values, calculated from the ratio of the matched-resolution dust continuum image to the integrated intensity line images as described in the text.  Fainter symbols have lower signal-to-noise ratio.  Black error bars show the median and mean absolute deviation in radial bins.  As in Figures \ref{scatterplot} through \ref{radialratios2}, these are deprojected radii.  A typical Milky Way value is also indicated 
\citep[e.g.][]{leroy_alphas}.
\label{fig:alphas}}
\end{figure}

Figure \ref{fig:alphas} shows the resolved CO-to-\htoo\ conversion factors $\alpha_X$ inferred for the three isotopologues under the assumptions detailed above.
Table \ref{tab:alphaX} also gives representative statistics for all of the pixels that 
have 3mm continuum emission detected at a signal-to-noise ratio $> 3,$
and that are free of synchrotron contamination (projected radius $r > 2.6$\arcsec, which is $1.3\times \mathrm{FWHM}$).
Our estimated \alphaCO\  value for \tweco(1$-$0) in NGC~4526 is 5.6$\pm$1.2 \msunpckkms, which is just 30\% larger than a ``standard" Milky Way value of 4.3 in the same units \citep[e.g.][]{leroy_alphas}.
The result suggests that \tweco\ emission in NGC~4526 is just a bit fainter than usual, for a given \htoo\ surface density.  Estimated values for \alphaCO\  under different assumptions may be scaled by noting that it is proportional to the gas/dust ratio and the fraction of the continuum intensity that is assumed to be dust, and that it is inversely proportional to the dust emissivity.
For example, if only half of the continuum comes from dust, our inferred value of $\alpha$ for \tweco(1$-$0) drops to 2.8$\pm$0.6 \msunpckkms.
As a consistency check, \citet{utomo4526} found that an \alphaCO\ value of 4.4 \msunpckkms\ suggests that most of the molecular clouds in NGC~4526 are in virial equilibrium.

Furthermore, as the RADEX/MCMC analysis gives confidence intervals on the species' column densities,
the \htoo\ column density can be used to infer the species' abundances.
In a continuum image at 99.33 GHz and 2.0\arcsec\ resolution, the surface brightness at $r=5$\arcsec\ is 0.12 $\pm$ 0.01 \mjb.  
Using the dust emissivity and gas/dust ratio described above, this specific intensity corresponds to an \htoo\ column density of 1.4\e{22}\persqcm, or 300 \solmass\persqpc\ (including helium).  Both of those values have been deprojected to face-on column densities assuming an inclination of 78$\arcdeg$ (Section \ref{sec:kin}).
The dust-derived \htoo\ column densities then give the CO isotopologue
abundances listed in the last column of Table \ref{tab:alphaX}.
For \tweco, the inferred abundance ranges from [\tweco/\htoo] $\approx$ 8\e{-6} to 3\e{-5}, roughly factors of 3 to 10 lower than the ``standard" Milky Way abundance of 8.5\e{-5} \citep{frerking82}.
The inferred \tweco\ abundance is proportional to the dust emissivity but is inversely proportional to the gas/dust ratio and to the fraction of the continuum that is assumed to be dust.  Thus, for example, if only half of the 3mm continuum emission came from dust, the \tweco\ abundance would be 
$\approx$ 2\e{-5} to 6\e{-5} and would be
closer to the Milky Way value.

Our quoted \tweco\ abundance in NGC~4526 is farther away from the ``standard" Milky Way value than one might expect, given that \alphaCO\  is fairly close to the Milky Way value.  
In this context it is worthwhile to note that the \tweco\ abundance calculation takes optical depths into account, because it is based on the RADEX/MCMC results.  In contrast, the \alphaCO\  calculation uses the raw line intensities and does not account for variations in optical depth.
Lower-than-usual \tweco\ optical depths in NGC~4526 might explain relatively normal \alphaCO\  values despite low \tweco\ abundances.

We do not place much significance on the differences between the \alphaCO\  and [\tweco/\htoo] values for NGC~4526 and the Milky Way, because
the uncertainties associated with this analysis are very large.
In Section \ref{sec:cont} we suggest that a free-free contribution at 100 GHz might mean that the dust surface brightnesses are half of the observed intensities.
Dust emissivities are also highly uncertain at these wavelengths \citep{cigan_sn1987a}; 
the slightly different modified black body parameters of \citet{auld2013} produce a difference of a factor of 3 in the inferred $\alpha_X$ values and the \htoo\ column density, and \citet{clark_dust} also find evidence for variations in the dust emissivity within an individual galaxy.
The RADEX/MCMC column densities are also, strictly speaking, not as well constrained as the ratios N$/\Delta v,$ and we have used line widths $\Delta v$ estimated from the outer part of the disk where there is little kinematic beam-smearing affecting the line widths.
Finally, estimates of gas/dust mass ratios find wide variations of at least a factor of 30, and sometimes 100, between different galaxies \citep{cortese2016,lianou}, and it is difficult to know how much of that variation is real and how much is noise.  Recent theoretical work suggests that much of it may be real \citep{whitaker}.

In the context of these uncertainties, we simply comment that the inferred $\alpha_X$ values and CO isotopologue abundances found here are roughly consistent with the values assumed in Section \ref{sec:kin}, and they are  similar to the spirals of the Local Group \citep{leroy_alphas}, and at our 150 pc resolution there is little to no evidence for strong internal variations within the disk of NGC~4526.

\begin{deluxetable}{lccc}
\tablecaption{Estimated CO/\htoo\ conversion factors and abundances\label{tab:alphaX}}
\tablehead{
\colhead{Species} & \colhead{$\alpha_X$} & \colhead{log N} & \colhead{[X/\htoo]} \\
\colhead{} & \colhead{} & \colhead{(cm$^{-2}$)} &\colhead{} 
}
\startdata
\tweco & 5.6$\pm$1.2 & 17.7 --- 18.2 &  7.7($-6$) -- 2.6($-5$) \\
\thirco & 19.3$\pm$4.1 & 16.9 --- 17.3 & 1.1($-6$) -- 2.8($-6$) \\
\ceighto & 108$\pm$29 & 16.1 --- 16.5 & 1.9($-7$) -- 4.4($-7$) \\
\enddata
\tablecomments{The values of $\alpha_X$ are quoted for the (1-0) transitions in units of \msunpckkms.
They represent the medians and mean absolute deviations of the selected pixels, as described in the text.  Column densities $N$ are 68\% confidence intervals from the RADEX/MCMC analysis (Section \ref{quantitative_ratios} and Appendix \ref{radex}) for conditions in the molecular ring, and they have not been deprojected to face-on values.
The abundances [X/\htoo] in the last column are also ranges propagated from the column density intervals, and 
the value in parentheses is the exponent.}
\end{deluxetable}

\section{Discussion}\label{sec:disc}

\subsection{Whole-galaxy integrated properties}\label{sec:disc_global}

Most of the molecular line ratios we have measured in NGC~4526, along with the inferred molecular abundances and CO/\htoo\ conversion factors, are similar to those found in nearby spirals.
In that broad sense, the molecular gas properties of NGC~4526 are consistent with the idea that at some time before it entered the Virgo Cluster it may have looked very much like a spiral galaxy.
(In this discussion we make the underlying assumption that it didn't acquire gas after entering the cluster, so whatever gas it contains now is gas that it either carried into the cluster upon entry or regenerated internally from stellar mass loss.)

One striking exception to the above analogy with spirals is the very low \tweco(1$-$0)/\thirco(1$-$0) line ratio
and the low inferred abundance ratios [\tweco/\thirco] = $7.8^{+2.7}_{-1.5}$ and $6.5^{+3.0}_{-1.3}$.
For comparison, estimated [\twec/\thirc] abundances in nearby spirals and the Milky Way tend to range from 20 to 60 \citep[e.g.][]{romano2019,tang2019,martin2019,alchemi1,teng_radex}.
In this regard, NGC~4526's status as a longtime, virialized resident of the Virgo Cluster is particularly relevant.
\citet{crocker_hd} and \citet{a3d_13co} noticed that Virgo Cluster early-type galaxies have systematically lower \tweco/\thirco\ line ratios than field early-type galaxies or spirals;
\citet{crocker_hd} suggested these low ratios might be caused by ram pressure stripping, which could remove a diffuse, optically thin component of the molecular gas.  In the case of NGC~4526, the \hi\ deficit (Section \ref{sec:about}) is also consistent with such stripping.
Our analysis confirms that the \thirco(1$-$0) emission is quite thick, with $\tau_{13} = 1.1^{+0.9}_{-0.5}$.  It is not yet clear whether the low abundance ratio [\tweco/\thirco] $\approx 7$ was established before or after NGC~4526 entered the cluster.
It would be interesting to compare that ratio to the expected nucleosynthetic yields, as NGC~4526 has much older stellar populations than the star-forming galaxies that commonly feature in isotopic analyses.

In contrast to \tweco/\thirco, the \thirco/\ceighto\ line ratios and abundances in NGC~4526 are similar to those found in spirals.  \citet{zhang2018} recently suggested that \thirco/\ceighto\ line ratios might be used as indicators of the initial mass function (IMF), as \eighto\ originates in more massive stars than \thirc, and both isotopologues are expected to have optically thin transitions.  Our analysis of NGC~4526 offers a cautionary note, since  the optical depth in \thirco(1$-$0) might be as large as 2.0.  \citet{romano2019} also note that uncertainties in nucleosynthetic yields, especially for \eighto, make it difficult to arrive at firm conclusions regarding the IMF.  Qualitatively and generally speaking, though, we would expect that recent star formation activity should drive down the [\thirco/\ceighto] ratio, which should then creep back upwards over several Gyr.   This general idea seems to suggest that NGC~4526 ought to have higher [\thirco/\ceighto] ratios than spirals, as it has a lower specific star formation rate, but quantitative models will be needed for a testable hypothesis.

\subsection{Structures within NGC~4526}\label{sec:disc_radial}

Outside of the nucleus, the most prominent structure in the molecular gas distribution of NGC~4526 is the ring at $r \sim 5.3$\arcsec\ (400 pc); the peak \tweco(1$-$0) and (2$-$1) intensities in the ring match those in the galaxy nucleus.  Here we refer to the region interior to the ring as the inner disk and to the exterior as the outer disk.
Until further kinematic studies are done, the origin of the ring and its possible dynamical significance are not clear.  However, we note that the most prominent star formation activity in NGC~4526, as traced by blue stars (Figure \ref{fig:mom0}) and low \oiiihb\ ratios (Figure \ref{fig:oiiihb}), is found just outside the molecular ring.
Figure \ref{scatterplot2} shows that the CN emission is optically thick in the inner disk and thin in the outer disk, but 
at present it is not clear whether the change in optical depth is directly driven by the star formation activity or whether it is unrelated.  Optical depth estimates in other high-density species would also be interesting in this regard.

The resolved data in this paper also show strong radial gradients in all of the line ratios between \tweco\ and other non-CO species.  In this regard, there doesn't appear to be a qualitative difference between the inner disk and the outer disk. 
However, there may be a qualitative difference between the inner disk and the outer disk in terms of the CO isotopologues.
Figure \ref{scatterplot} and Appendix \ref{radex} suggest that \thirco\ and \ceighto\ are well mixed in the inner disk, and the majority of the radial variation (a factor of 1.6 in abundance, or 1.8 in the line ratio) occurs only in the outer disk at $r > 400$ pc.  In this subsection we discuss some of the plausible interpretations of the [\thirco/\ceighto] abundance gradient and the different behaviors in the inner and outer disks.

For comparison, \citet{jd_13co} have presented data on resolved \thirco/\ceighto\ line ratios in a small handful of nearby spiral galaxies.  They analyze line ratios, not abundances, but they argue their lines are probably optically thin so that the line ratios are good proxies for abundances.  Most of their nearby spirals show large-scale radial gradients in \thirco/\ceighto, with generally rising profiles that have values of 4 to 6 in their centers and values of 10 to 20 at $\approx$ 6 kpc.
On smaller scales, their ALMA data sometimes show strong local variations within the inner 3 kpc; two interesting cases show \thirco/\ceighto\ rising by a factor of two and then falling again, producing a local maximum at about 700 pc.  Clearly, nearby galaxies show a variety of behaviors in these isotopic line ratios, and the line ratios should be sensitive diagnostics of conditions within their disks.

Isotope-selective photodissociation is frequently mentioned as a possibility for explaining spatial variations in isotopic line ratios \citep[e.g.][]{rodriguezbaras}.  \citet{jd_13co} dismiss it in their cases on quantitative grounds.  In NGC~4526, the sense of the [\thirco/\ceighto] abundance gradient is such that explaining it with isotopic-selective photodissociation would require stronger UV fields in the outer part of the molecular disk.  That might be plausible in a spiral whose colors become increasingly bluer at large radii, but it seems unlikely for an early-type galaxy.  
Furthermore, isotope-selective photodissociation should also produce an increase in [\tweco/\thirco] with radius, and we do not observe that effect.  

Another astrochemical process that can affect isotopologue abundances is fractionation \citep[e.g.][]{viti2020}; in regions where the temperature is fairly low but C$^+$ is still available for ion-molecule reactions, the CO molecules can be preferentially enhanced in \thirco\ relative to \tweco.   \citet{langer84} find the process is important for $T \lesssim 35$ K, and the higher ionization potential of oxygen means this process affects the C isotopes but not the O isotopes.  Thus the effect would be expected to produce mirror-image radial trends in [\tweco/\thirco] and [\thirco/\ceighto]; if the one increases with radius, the other should decrease.  
\citet{jd_13co} favor this process as an explanation for the large-scale rising \thirco/\ceighto\ line ratios they observe over several kpc in spiral galaxy disks, but they do not yet have the corresponding [\tweco/\thirco] abundance estimates to test the hypothesis.
In the case of NGC~4526, our temperature estimates from CO, HNCO and [\ion{C}{1}] are all low enough to make the fractionation process relevant.  However, the dramatic radial increase in [\thirco/\ceighto] is not mirrored by a similarly strong decrease in [\tweco/\thirco].
Thus it also seems unlikely that fractionation alone drives the [\thirco/\ceighto] gradient in NGC~4526.

It is also possible that the steep radial gradient in [\thirco/\ceighto] in the outer disk of NGC~4526 reflects a compression effect driven by the galaxy's presence in the Virgo Cluster.
Specifically, \citet{tonnesen2009} note that hydrodynamical interactions between a spinning galaxy disk and the non-rotating intracluster medium (ICM) must inevitably result in angular momentum transfer from the rotating gas disk to the ICM, causing the disk gas to gradually spiral inward.  In their simulated case with the highest ICM thermal pressure, this effect makes the outer edge of gas disk creep inward from 20 kpc to 12 kpc over 1 Gyr, even in the absence of any ram pressure stripping.  If the gas annuli creep inward at similar radial velocities, this process could compress the gas disk while maintaining pre-existing abundance gradients.  It is interesting to speculate, therefore, whether the [\thirco/\ceighto] gradient in the disk of NGC~4526 represents several kpc worth of typical spiral galaxy abundance pattern gradients, radially compressed into the current 1 kpc disk.
Any compression on that scale must also have been accompanied by ram pressure stripping, though, as the gas surface densities and volume densities are not unusual (Sections \ref{sec:kin} and \ref{sec:coradexresults}).
Furthermore, secular evolution within the disk of NGC~4526 has probably been complicated by the presence of a bar \citep{michard,ferrarese}; the bar has been identified based on boxy/peanut-shaped isophotes (Figure \ref{fig:4526_hst}), and it might have driven the formation of the molecular ring.

Finally, the radial gradient in [\thirco/\ceighto] in NGC~4526 might be driven by recent star formation activity in the inner disk, increasing the \eighto\ abundance there.
A detailed test of this idea would benefit from a higher-resolution image of an extinction-free or at least extinction-corrected star formation tracer, and from analysis of the gravitational instabilities in the molecular disk.
At the moment we simply note that the most obvious blue stars and low \oiiihb\ ratios are located not in the inner disk, where the [\thirco/\ceighto] abundance is lowest, but just exterior to the molecular ring.  The displacement is reminiscent of features sometimes seen in spiral arms, where the blue stars are displaced downstream of the arm ridgeline traced by molecular gas.  In general, early-type galaxies are not obvious targets for detailed studies of the star formation process.  However, NGC~4526's long residence in the Virgo Cluster (its interactions with the ICM) and its very deep internal potential well (its unusually large circular velocity) make it an interesting target for star formation and chemical evolution studies.

\section{Summary}

We present new ALMA observations of continuum emission and several molecular species in the dusty disk of the Virgo Cluster early-type galaxy NGC~4526; the species include CO and its isotopologues, CN, CS, HCN, \hcop, HNC, HNCO, and \chhhoh.

The 3mm continuum emission shows a nuclear point source with a spectral index of $-1.07 \pm 0.02,$ which is clearly synchrotron emission from an AGN that is also detected in X-rays and in lower frequency radio emission.
We also find low surface brightness continuum emission from an extended disk; this disk emission has a strongly rising spectral index, and its flux density is a remarkably good match to the long-wavelength extrapolation of a modified black body that fits the FIR flux densities.  We therefore estimate that 50\% to 100\% of this extended 3mm continuum is dust emission, and the remainder is probably free-free emission.

The resolved molecular disk structure shows a nuclear peak and a bright ring at $r \sim 5$\arcsec\ (0.4 kpc); assuming a CO/\htoo\ conversion factor of 4.3 \solmass\persqpc\ (K \kms)$^{-1}$, the deprojected molecular surface densities in the nucleus and in the ring are 240 \solmass\persqpc\ at 75 pc resolution.
The molecular disk also shows modest asymmetries, kinematic disturbances, and evidence for a bar, and those kinematic features will be discussed in a future paper.

The \tweco(1$-$0)/\thirco(1$-$0) line ratio in NGC~4526 is one of the lowest ever observed outside of the Local Group, at 3.5 $\pm$ 0.35, and it is remarkably constant throughout the disk.
In contrast, \thirco(1$-$0)/\ceighto(1$-$0) shows typical values but an unusually steep gradient; it is nearly constant within the molecular ring ($r < 5$\arcsec) but exterior to that it rises by almost a factor of two over only 0.6 kpc in radius.

We carry out RADEX modeling of the CO isotopologues and employ Bayesian analysis with a Markov chain Monte Carlo sampler to estimate posterior probability distributions for the physical parameters in the CO-emitting gas.
The current set of transitions strongly constrain the [\thirco/\ceighto] abundance ratio to be 5.7$^{+0.7}_{-0.6}$ in the inner disk ($r < 2$\arcsec)
 and 9.4$^{+1.1}_{-0.9}$ at a radius of 1 kpc;
 those uncertainties are 68\% confidence intervals.
The [\tweco/\thirco] abundance ratio is poorly constrained in the very inner regions of the disk but is found to be 
$7.8^{+2.7}_{-1.5}$ at $r \approx 5$\arcsec\ and $6.5^{+3.0}_{-1.3}$ at $r \approx 12$\arcsec, and these values are a factor of three lower than those typically found in nearby spirals.
 
 We find volume densities \nhtoo\ = 250\percc\ $\pm$ 0.4 dex and \tkin\ $> 10$ K in the CO-emitting gas.
The \thirco(1$-$0) line is moderately optically thick, with $\tau_{13} \approx 1$ and possibly as high as 2. 
Our temperature estimates from the CO lines are consistent with those from the FIR [\ion{C}{1}] transitions and from CN (see below).  We also infer temperatures in the range of 5.4 K to 13.4 K from HNCO(5$-$4)/(4$-$3).  These estimates  roughly follow the expected pattern that the high-density tracers should show lower temperatures than CO or [\ion{C}{1}], reflecting the origins of the high-density tracers in the darker and colder interiors of molecular clouds.  The uniformly low temperatures are also consistent with general expectations based on the galaxy's low star formation rate.

NGC~4526 is relatively bright in the high-density tracers HCN, CN, and especially CS; it is relatively faint in the shock tracers \chhhoh\ and HNCO.
We find no measurable enhancements in high-density ratios like CN/HCN or HCN/\hcop\ (or indeed anything else) towards the AGN, when measured at 150 pc resolution.  The one possible exception to that statement is a tentative rise in \tweco(2$-$1)/(1$-$0) towards the nucleus, but that analysis is complicated by uncertain continuum subtraction in the (2$-$1) data.
It has also been suggested that CS/HCN might be enhanced in regions of star formation activity, but we find no evidence for any spatial variations in that ratio either.  Line ratios between any two high-density tracers are remarkably constant within NGC~4526.

NGC~4526 also has an unusually high optical depth in the 3mm CN  (N=1$-$0) lines.
We find an extremely low (J=3/2$-$1/2)/(J=1/2$-$1/2) ratio of 1.12 $\pm$ 0.16 in the nucleus, rising to values consistent with 2.0 outside the molecular ring ($r > 5$\arcsec).
An assumption of LTE requires temperatures $\leq$ 20 K in order to reproduce the low line ratio in the nucleus, and it suggests the optical depth in the 3/2$-$1/2 transition is $\tau_{3/2} \geq 4.5.$  
For typical abundance ratios, this value then suggests that HCN(1$-$0), \hcop(1$-$0), and HNC(1$-$0) should also be very thick, with $\tau_\mathrm{HCN 1-0}$ in the range of 5 to 30.
Extragalactic observations rarely find such high values, and it is not clear why NGC~4526 would be unusual in this respect. 

Motivated by the good match between the modified black body fit and the extended 3mm continuum flux density, we assume a dust emissivity and a gas/dust ratio in order to infer dust-based \htoo\ surface densities and CO/\htoo\ conversion factors \alphaCO.  Column density estimates for the CO isotopologues, from the RADEX/MCMC analysis, then allow us to estimate the abundances of the CO species.  The values have large uncertainties of at least a factor of a few, but at that level they are consistent with other values inferred for nearby spirals.  We find no particular evidence for internal variations in \alphaCO\  or the gas/dust ratio, though we cannot estimate those quantities toward the nucleus because of the dominant synchrotron continuum component there.

In many respects, the molecular properties of NGC~4526 are consistent with it being a heavily stripped descendant of something that used to look like a large spiral galaxy.  Its unusually low [\tweco/\thirco] abundance ratios and typical [\thirco/\ceighto] ratios provoke questions about the processing of its molecular gas in the Virgo Cluster.
Its deep potential well (high circular velocity) should protect its molecular gas from stripping more effectively than would be the case in lower-mass spirals, though, so it's not obvious whether the galaxies that are currently being stripped in the Virgo Cluster will end up looking like NGC~4526 in the future.
Future work on isotopologues in other cluster galaxies, in combination with chemical enrichment models, will be particularly valuable for understanding the history and the processing of galaxies in clusters.

\acknowledgments

We thank Martin Bureau for many fascinating and productive conversations about early-type galaxies.
This paper makes use of the following ALMA data: ADS/JAO.ALMA\#2017.1.01108.S and \\
ADS/JAO.ALMA\#2018.1.01599.S.
ALMA is a partnership of ESO (representing its member states), NSF (USA) and NINS (Japan), together with NRC (Canada), MOST and ASIAA (Taiwan), and KASI (Republic of Korea), in cooperation with the Republic of Chile. The Joint ALMA Observatory is operated by ESO, AUI/NRAO and NAOJ.
The National Radio Astronomy Observatory is a facility of the National Science Foundation operated under cooperative agreement by Associated Universities, Inc.

\facilities{ALMA, CARMA, IRAM}

\appendix

\section{Verification of \tweco (2-1) data}\label{carmadata}

\begin{figure}
\includegraphics[width=3.5in,trim=3mm 3mm 1cm 0cm,clip]{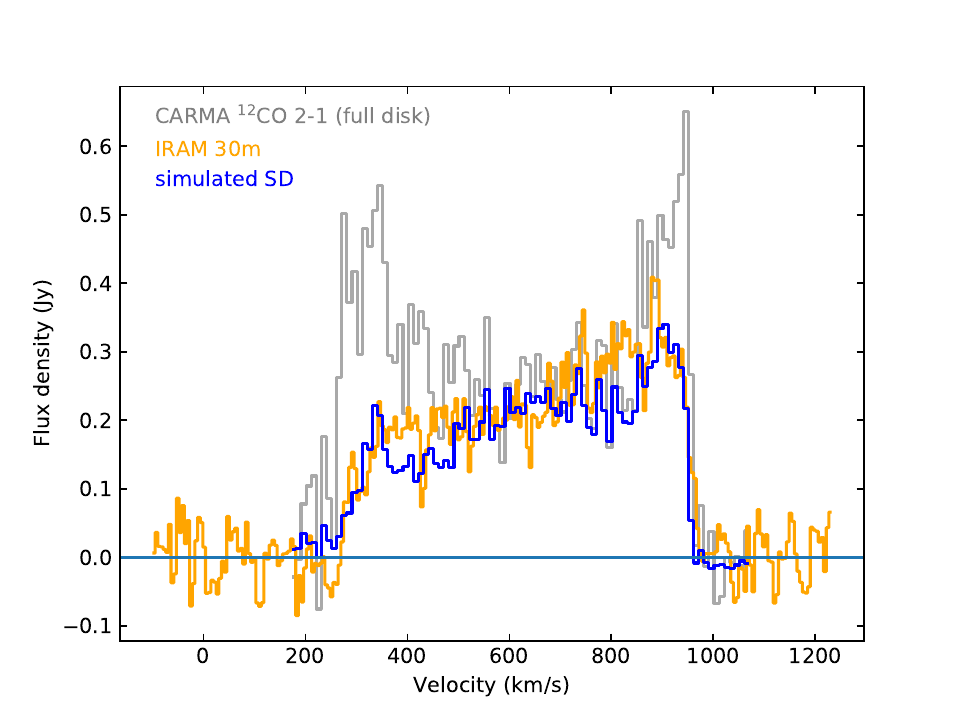}
\caption{\tweco(2-1) spectra of NGC~4526.  The original CARMA spectrum is integrated over the full region of emission as defined by \tweco(1-0) and \thirco(1-0) detections in the ALMA data.  We show also the original IRAM 30m \tweco(2-1) spectrum and a simulated version of a single-dish spectrum recreated from the CARMA data cube by convolving and applying a modest pointing offset.\label{21spec}}
\end{figure}

The \tweco(2-1) data used here were obtained with the CARMA telescope at a resolution of 0\arcsec.25 by \citet{davis_nature}.  A smoothed version of the data was also used in the dynamical analysis of \citet{davis+mcd_imf}.  As the {\it uv}-coverage of the data is significantly different from that of the other lines studied here, we check whether the CARMA images might have missed significant flux by comparing them to a single-dish spectrum from the IRAM 30m telescope \citep{CYB}.
We convolve the CARMA data cube to the 12\arcsec\ resolution of the 30m at 230 GHz, ensuring proper flux retention by appropriate normalization of the convolution kernel.
The spectra of the individual pixels in the convolved cube then simulate what the 30m would have observed if it were pointed at that location.  The technique account can account for both pointing offsets and the fact that the 30m beam is smaller than the size of the molecular disk.
The results are shown in Figure \ref{21spec} for a location 1.6\arcsec\ northwest of the galaxy nucleus, where the location is chosen by eye to reproduce the degree of asymmetry in the 30m spectrum.  (Larger offsets produce stronger asymmetry and smaller offsets produce weaker asymmetry.)   An offset of 1.6\arcsec\ is well within the expected pointing performance of the 30m telescope.\footnote{https://www.iram-institute.org/EN/content-page-58-7-55-58-0-0.html}  The good match achievable between the actual 30m spectrum and the simulated one suggests the CARMA \tweco(2-1) data do not miss significant flux due to missing short spacings.  Indeed, in a source like this with such strong rotation, the emission in each individual channel is effectively unresolved in the direction of the local velocity gradient.

A second issue for the \tweco(2-1) images is that continuum subtraction was not performed on the original data because it was primarily used for kinematic work, and the limited available bandwidth did not provide enough sensitivity in line-free channels to detect any continuum emission.
But we can now estimate the strength of the 230 GHz continuum, based on the 100 GHz continuum and FIR flux density measurements. 
As described in Section \ref{sec:cont}, the 99 GHz continuum image at 2.0\arcsec\ resolution (Figure \ref{fig:dust+cont}) contains roughly 1.9$\pm$0.1 mJy in an extended disk with a spectral index of +1.8 and 4.7$\pm$0.1 mJy in a nuclear point source with a spectral index of $-1$.
Uncertainties in the way the flux density is partitioned are difficult to estimate because they involve interpolating the disk under the point source, so the quoted uncertainties may well be underestimates.
Extrapolating those flux densities using their estimated spectral indices suggests 11 mJy of total continuum emission at 230 GHz, of which roughly 2 to 4 mJy would be the nucleus and the rest would be the disk.
Alternatively, the modified black body fits, which use a steeper effective spectral index for the disk emission, predict roughly 40 mJy at 230 GHz (Figure \ref{fig:4526sed}).

We construct a model 230 GHz continuum image including 30 mJy in an elliptical Gaussian mimicking the size and shape of the dust disk, plus 4 mJy in a nuclear point source.  This level of continuum emission is below the detection limits of the CARMA data.  We then subtract this model continuum image from the CARMA data cube and compare analyses with and without the estimated continuum.
The continuum subtraction makes a 10\% difference in the integrated \tweco(2-1) flux (221 $\pm$ 6 Jy~\kms\ compared to 245 $\pm$ 6 Jy~\kms).
It also makes a 35\% difference in the nuclear peak integrated intensity (8.1 $\pm$ 0.7 \jybkms\ at 1.12\arcsec\ resolution, vs 11.0 $\pm$ 0.7 \jybkms).
Resolved line ratio measurements outside of the nucleus are indistinguishable in the two cases, as the disk has low surface brightness and the line emission outside of the nucleus has a relatively narrow velocity support.
The values presented in the paper are computed after the estimated continuum has been subtracted.

\section{RADEX + MCMC analysis}\label{radex}

We investigate the physical conditions in the CO-emitting gas through the RADEX non-LTE code, which carries out iterative solutions of the level populations and the line radiation using an escape probability formalism \citep{radex}.  
We compute the line intensities and optical depths for the $J=1-0$ and $2-1$ transitions in \tweco,\thirco, and \ceighto\ for a grid of physical conditions, giving the relevant line ratios as functions of those physical conditions.
The grid spans volume densities \nhtoo\ = 50 to 10$^6$\percc\ in steps of 0.215 dex, kinetic temperature \tkin\ = 3.0 to 80 K in steps of 0.071 dex, and column densities N(\tweco) = 8\e{15} to 8\e{19}\persqcm\ in steps of 0.067 dex.  
For a fiducial [\tweco/\htoo] abundance of 8\e{-5} \citep[e.g.][]{frerking82}, the corresponding total column densities are  N(\htoo) = 1\e{20} to 1\e{24}\persqcm. 
We also assume a fixed linewidth of 30 \kms, which is similar to the linewidth in the outer parts of the galaxy where there is little beam smearing (Figure \ref{fig:veldisp}).  In this context we note that the RADEX calculations in fact depend only on the ratio of the column density to the line width, so we show fiducial column densities N(\tweco) in the figures but in fact the fundamental parameter of physical relevance is N(\tweco)/$\Delta v$.  The corresponding column densities can be easily calculated for other linewidths with a simple scaling.  The grid 
also spans [\tweco/\thirco] abundance ratios from 3.4 to 400 and [\thirco/\ceighto] from 2.0 to 27, both in steps of 0.067 dex.  This grid strategy is very similar to that employed by \citet{teng_radex}, though independently coded.

We then compare the predicted line ratios to observed quantities and use a Markov chain Monte Carlo sampler \citep[emcee;][]{emcee} to estimate posterior probability distributions on the relevant physical parameters in the molecular gas.
Sampling using the adaptive Metropolis algorithm of \citet{cap:a3dJAM} produces results that are indistinguishable from those of {\it emcee}.
In both cases, the sampler interpolates in the RADEX grids to evaluate predicted intensities, line ratios, and the likelihood at each set of physical conditions it explores.  We assume flat priors on the logs of the parameters, over the range of conditions spanned by the RADEX grids.

The observations we use as constraints are the \tweco(1$-$0) intensity and the 
line ratios \tweco(1$-$0)/\thirco(1$-$0), \tweco(1$-$0)/\ceighto(1$-$0), \thirco(1$-$0)/\ceighto(1$-$0), \tweco(2$-$1)/\tweco(1$-$0), and \tweco(2$-$1)/\thirco(2$-$1).
As the beam is larger than the typical size of a molecular cloud, beam dilution may make the observed \tweco(1$-$0) brightness temperature smaller than the intrinsic radiation temperature for the line.  We account for the beam filling factor by using an asymmetric distribution for the \tb\ contribution to the model likelihood.  Specifically, the other measurements' contributions to the model likelihood are symmetric Gaussians, with their measurement uncertainties characterizing the width of the distributions, but for \tb\ we use a Gaussian below the measured value and a flat distribution above the measured value.  All predicted line intensities that are {\it larger} than the measured value are thus considered equally likely, allowing for beam filling factors in the range $0 \le f \le 1.$  In attempting to reproduce the other line ratios we are effectively assuming this beam filling factor is the same for all the isotopologues.

The \tweco(2$-$1)/\thirco(2$-$1) line ratio also deserves some comment; it is not measured using resolved ALMA data, like the other line ratios, but with the IRAM 30m telescope \citep{crocker_hd}
so it is an average over a 12\arcsec\ beam.  The fact that the resolved measurements of \tweco(1$-$0)/\thirco(1$-$0) and \tweco(2$-$1)/\tweco(1$-$0) show no significant radial variation lends confidence to the assumption that 
\tweco(2$-$1)/\thirco(2$-$1) may also be roughly constant over the disk of NGC~4526.

We carry out this analysis for three representative sets of conditions in NGC~4526.  One set of values corresponds to the nuclear regions ($r \lesssim 2$\arcsec), where the \ceighto\ intensity is relatively high, so the \thirco/\ceighto\ intensity ratio is low, and the \tweco(1$-$0) brightness temperature is lower than elsewhere in the galaxy (Section \ref{sec:kin}).
Another set corresponds to values measured in the molecular ring at $r \approx 5$\arcsec, where the \thirco/\ceighto\ intensity ratio is still low like the inner disk but the \tweco(1$-$0) brightness temperature is at its highest.  The third set of values corresponds to the outer disk, where the \thirco/\ceighto\ intensity ratio is relatively large and \tweco(1$-$0) brightness temperatures are intermediate between those of the nucleus and the ring.

\begin{figure}
\includegraphics[width=\textwidth, trim=0mm 5mm 1cm 0cm,clip]{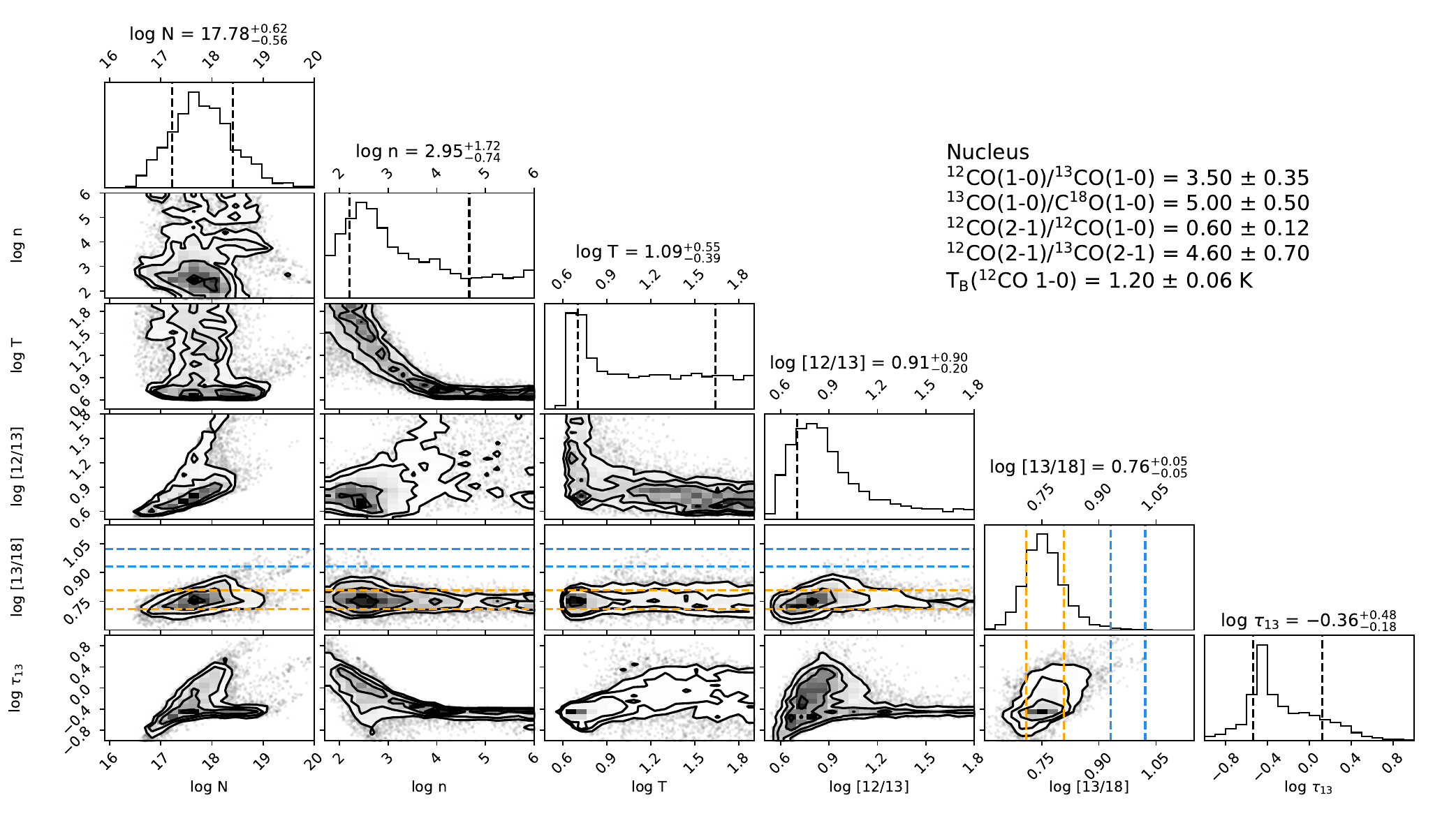}
\caption{Corner plot \citep{corner} for the physical conditions reproducing the line ratios observed in the nuclear regions of NGC~4526, $r \lesssim 2$\arcsec.  The greyscale intensities and corresponding marginalized histograms show the density of samples obtained by the MCMC sampler.  The label ``log N" refers to the log of the \tweco\ column density in\persqcm, for an assumed linewidth of 30 \kms\ FWHM; 
``log n" refers to the \htoo\ density in\percc; T is the kinetic temperature in K; ``[12/13]" is the [\tweco/\thirco] abundance ratio; ``[13/18]" is the [\thirco/\ceighto] abundance ratio; and $\tau_{13}$ is the optical depth of the \thirco(1-0) line.
Column densities have not been deprojected to face-on values.
The optical depth is, of course, a dependent parameter rather than an independent one, and it is shown for reference because it guides our interpretations of the line ratios.
Above the one-dimensional histograms, titles give the medians of the samples of each parameter and the uncertainties quoted indicate the range between the 16\% and 84\% quantiles in the cumulative distributions (dashed black lines).  In the panels showing [\thirco/\ceighto], orange dashed lines indicate the values derived for the nuclear regions and blue dashed lines show the values for the outer edge of the disk.  The measured parameters used for this analysis are shown in the upper right corner of the figure.
\label{corner_inner}}
\end{figure}

\begin{figure}
\includegraphics[width=\textwidth, trim=0mm 5mm 1cm 0cm,clip]{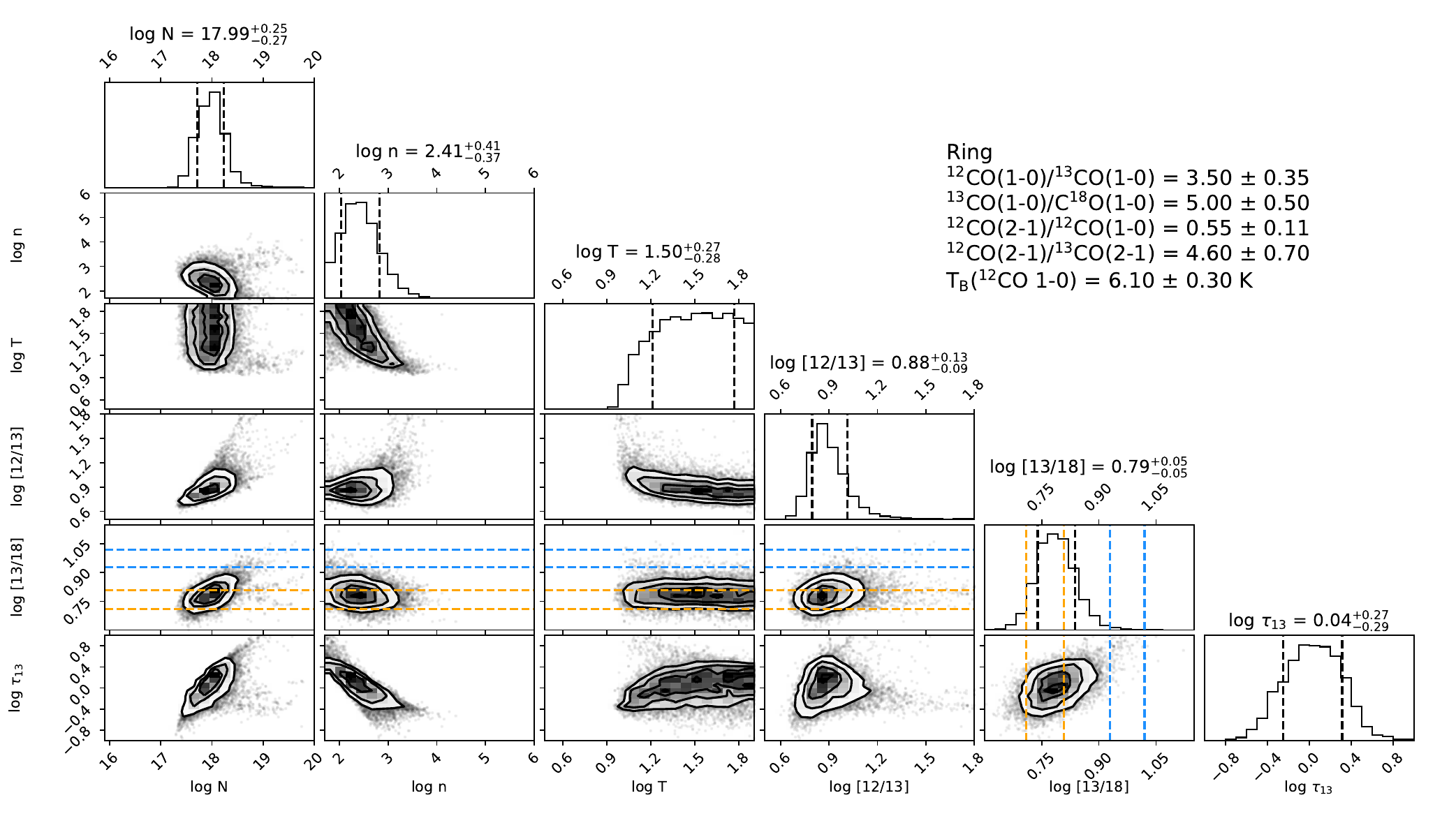}
\caption{Corner plot for the conditions in the molecular ring, at $r \approx 5$\arcsec.  Labels and lines have the same meanings as in Figure \ref{corner_inner}.\label{corner_ring}}
\end{figure}

\begin{figure}
\includegraphics[width=\textwidth, trim=0mm 5mm 1cm 0cm,clip]{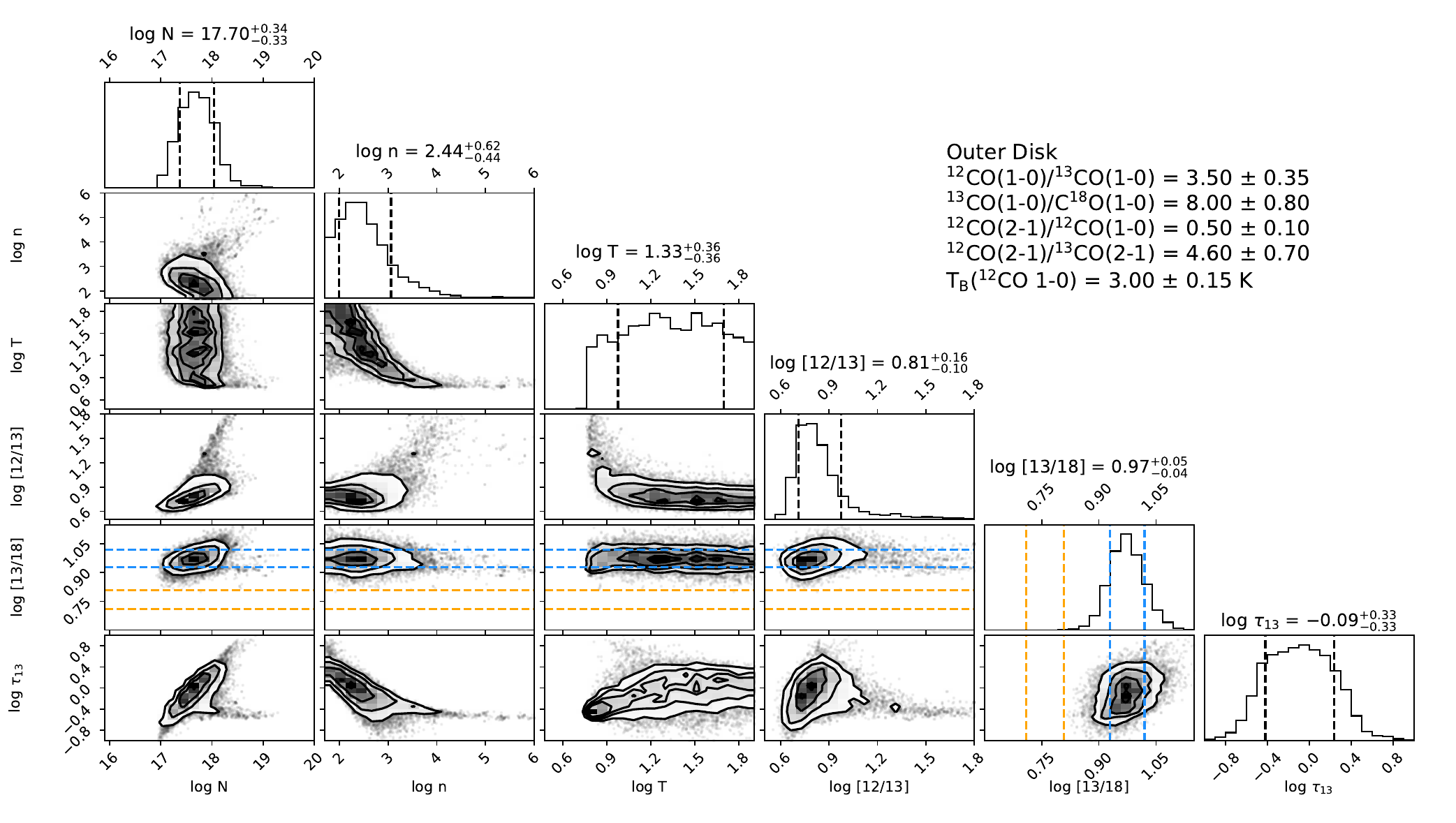}
\caption{Corner plot for the outer molecular disk of NGC~4526.  Labels and lines have the same meanings as in Figure \ref{corner_inner}.\label{corner_outer}}
\end{figure}

Figures  \ref{corner_inner}, \ref{corner_ring}, and \ref{corner_outer} show corner plots for the posterior probability distributions for these conditions.
Figure \ref{corner_inner} is for the nucleus; it illustrates the classic degeneracy between density and kinetic temperature for driving excitation, due to the limited set of currently available line ratios and the fact that the \tweco\ brightness temperature does not provide strong constraints in the nucleus.
The inferred posterior probability distribution for the [\tweco/\thirco] abundance ratio peaks at 6.3 in the nucleus, though there is a long asymmetric tail such that higher values are also permitted.  The inferred [\thirco/\ceighto] abundance ratio is, however, tightly constrained and almost independent of the other parameters; it is $5.7^{+0.7}_{-0.6}$ (the quoted uncertainties correspond to a 68\% confidence interval).  That value is 15\% larger than the observed line ratio, due to the optical depth of the \thirco(1$-$0) line.
 
The conditions in the ring (Figure \ref{corner_ring}) include higher brightness temperatures, which give tighter constraints on the excitation.  Here the line ratios suggest relatively low densities for the molecular gas, with \nhtoo\ $\sim$ 250\percc\ $\pm$ 0.4 dex and \tkin\ $> 10$ K.   The improved constraints on \nhtoo\ and \tkin\ also help to constrain the [\tweco/\thirco] abundance ratio, where we find  $7.8^{+2.7}_{-1.5}$.
We also find a [\thirco/\ceighto] abundance of $6.2^{+0.8}_{-0.6}$, consistent with the value in the nucleus.  
These conditions require \thirco(1$-$0) to be moderately optically thick, with $\tau_{13} = 1.1^{+0.9}_{-0.5}$.
As a check on the plausibility of those conditions, we note that the density \nhtoo, column density N(\tweco), and a typical CO abundance of $10^{-4}$ give a line-of-sight thickness of 12 pc.  That value is reasonable for a giant molecular cloud; additional discussion can be found in \citet{teng_radex}.
Finally, in the outer disk (Figure \ref{corner_outer}) we infer similar \nhtoo, \tkin, and [\tweco/\thirco] values to the ring, with [\tweco/\thirco] =  $6.5^{+3.0}_{-1.3},$
but substantially higher [\thirco/\ceighto] abundances of $9.4^{+1.1}_{-0.9}.$

In short, we find strong evidence for a radial gradient in the [\thirco/\ceighto] abundance ratio, increasing to larger values at larger radii; the change is a 
factor of 1.6 over 1 kpc.  The gradient is shallower in the inner disk and steeper in the outer disk, with the majority or perhaps all of the gradient occurring exterior to the ring.
In contrast, we find no compelling evidence for a radial gradient in the [\tweco/\thirco] ratio, as our two inferred values of $7.8^{+2.7}_{-1.5}$ and $6.5^{+3.0}_{-1.3}$ are consistent with each other given their uncertainties.
However, those [\tweco/\thirco] abundance ratios are unusually low.  If they reflect an intrinsic [\twec/\thirc] ratio $\approx 7$,  then this isotopologue ratio in NGC~4526 is at least a factor of three lower than the centers of most nearby spirals \citep[e.g.][]{martin2019}.  Additional discussion of the implications of these isotopologue measurements is in Section \ref{sec:disc}.

\begin{figure}
\includegraphics[width=0.5\textwidth, trim=1cm 3mm 1cm 1cm,clip]{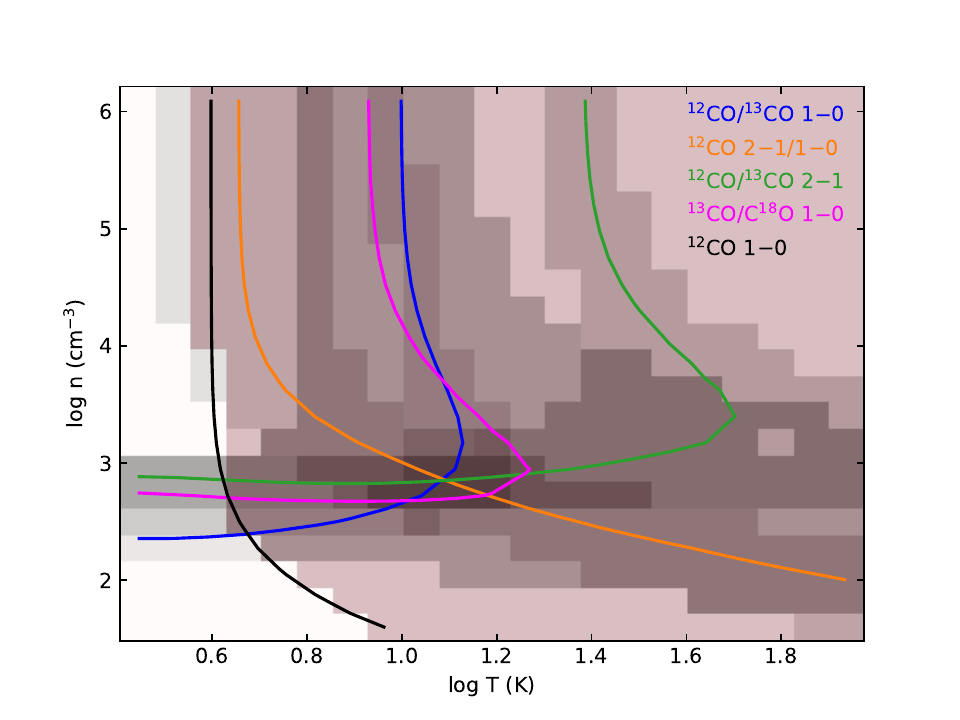}
\includegraphics[width=0.5\textwidth, trim=1cm 3mm 1cm 1cm,clip]{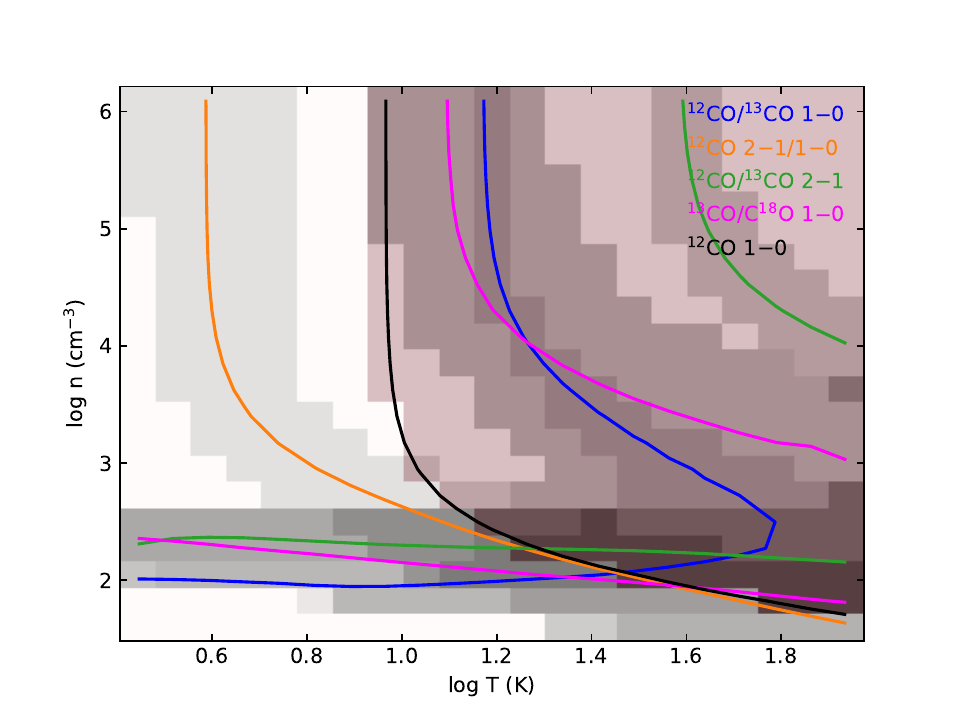}
\caption{Representative slices of the RADEX model volume, corresponding to the nucleus (left) and the molecular ring at $r \approx 5$\arcsec\ (right).
Contour lines mark the conditions that reproduce the measured values.  Gray bands around each contour line are within $\pm 1\sigma$ of the measured line ratios; for the \tweco(1$-$0) brightness temperature measurement, the pink shaded region is the allowed region with predicted brightness temperatures higher than the measured value and an inferred beam filling factor in the range $0 \leq f \leq 1$.  Darker pixels thus have higher likelihood.  Additional information is in the text.
\label{fig:nTsanitycheck}}
\end{figure}

Figure \ref{fig:nTsanitycheck} also presents (\nhtoo, \tkin) slices of the 5-dimensional RADEX model volume, with contours indicating the measurements we are attempting to explain.  
We construct each panel by identifying the most likely values for the \tweco\ column density, [\tweco/\thirco] and [\thirco/\ceighto] abundance ratios for each set of modeled conditions.  (Specifically, we use the median values of the marginalized posterior probability distributions, as indicated in the histogram titles in Figures \ref{corner_inner} and \ref{corner_ring}.)  The slice locations are fixed at the closest grid values to the medians in those three parameters.  As usual \citep[e.g.][]{teng_radex} the maxima in the individual one-dimensional marginalized distributions do not necessarily correspond to the maximum likelihood solution in the whole model volume, but they are close enough to be used as reasonable plausibility checks to illustrate the properties of our solutions.
For improved constraints on the temperature of the gas, and thus on its density, we suspect that it would be most helpful to have high resolution measurements of the \thirco(2$-$1) and \ceighto(2$-$1) lines.

\bibliography{alma4526lines}



\end{document}